\documentclass{aa}
\usepackage{txfonts}
\usepackage{amsmath}
\usepackage{amssymb}
\usepackage{multicol}
\usepackage{bm}
\usepackage{pdflscape}
\usepackage{ae,aecompl}
\usepackage{paralist}
\usepackage{lscape}
\usepackage{rotating}
\usepackage{afterpage}
\usepackage{tabularx}

\usepackage{natbib}

\begin{document}
\title{Populations of super-soft X--ray sources in galaxies of different morphological types}

\titlerunning{Populations of super-soft X--ray sources in galaxies of different morphological types}

\author{ I. Galiullin\inst{1,2} \and M. Gilfanov\inst{1,3} }

\institute{Max Planck Institute for Astrophysics, Karl-Schwarzchild-Str.1, Garching b. Munchen D-85741, Germany \email{ilkham@MPA-Garching.MPG.DE} \and Kazan Federal University, Kremlevskaya Str.18, 420008, Kazan, Russia \and Space Research Institute of Russian Academy of Sciences, Profsoyuznaya 84/32, 117997 Moscow, Russia }

\date{Received 24 September 2020; Accepted 30 November 2020}

  \abstract
   {}
   { We study populations of soft and super-soft X-ray sources in nearby galaxies of various morphological types with the special emphasis on characterizing  populations of stable nuclear burning accreting WDs.}
   {Analysing the content of Chandra archive we assembled  a sample of  nearby galaxies suitable for studying populations of super-soft X-ray sources.  Our sample includes 4 spiral galaxies, 2 lenticular galaxies and 3 ellipticals with stellar mass exceeding $10^{10}$ $M_\odot$ and X-ray sensitivity of the order of a  ${\rm few}\times 10^{36}$ erg/s. We used combination of hardness ratio and median energy to pre-select  X-ray sources with soft spectra, and  temperature  -- X-ray luminosity diagram to identify super-soft X-ray sources --  likely nuclear burning accreting white dwarfs.}
{For spiral galaxies, there is a distinct and rare population of    super-soft sources, largely  detached from the rest of sources on the $kT_{bb}-L_X$ plane. The boundary between these sources and the much more numerous population of harder (but still soft) sources is consistent with the  boundary of stable hydrogen burning on the white dwarf surface. Combined spectrum of soft sources located outside this boundary,  shows clear emission lines of Mg and S, which equivalent width is similar to that in the combined spectrum of a large number of confirmed supernova remnants in M83. This confirms earlier suggestions that the vast majority of so called  quasi-soft sources are supernova remnants. In early-type galaxies, populations of super-soft sources are about a factor of $\approx 8$  less abundant,  in broad agreement with the population synthesis calculations. Specific frequencies of super-soft sources  are:  (2.08$\pm$0.46)$\times10^{-10}$ M$_{\odot}^{-1}$  in spiral galaxies and  (2.47$\pm$1.34)$\times10^{-11}$ M$_{\odot}^{-1}$ in lenticular and elliptical galaxies, with the ratio of  the latter to the former of $0.12\pm0.05$. }
 {}
\maketitle

\section{Introduction}

CAL 83 and CAL 87 -- the two prototypical super-soft X-ray sources were discovered by Einstein observatory in the course of a survey of the Large Magellanic Clouds \citep{1981ApJ...248..925L}. About a decade later, the first light observations of central region of LMC with ROSAT observatory led to a discovery of a similar source with a very soft spectrum, RX J0527.8-6954 \citep{1991Natur.349..579T}. With advent of Chandra and XMM-Newton, such sources were found in  other, more distant, galaxies. Spectra of these sources are very soft and  do not extend beyond $\sim 1-2$ keV, their approximation with the black body model giving temperatures of the order of  $\sim 10-100$ eV and bolometric luminosities in the  $\sim 10^{36}-10^{38}$ erg/s range \citep{1991A&A...246L..17G, 1994A&A...288..538K}. It is the extreme softness of their spectra that gave rise to their name  "super-soft X-ray sources" (SSS). Binary nature of some of these sources (CAL 83 and CAL 87 in the first place) was determined soon after their discovery \citep{1988A&A...203L..27P,1988MNRAS.233...51S,1990ApJ...350..288C} and, although they were initially thought to be low-mass X-ray binaries, it was soon realised that the compact object in these binaries is a  white dwarf rather than a neutron star or a black hole \citep{1992A&A...262...97V}. The proposition that the main source of energy in super-soft X-ray sources is thermonuclear burning of the accreted hydrogen naturally explained the unusual softness of their X-ray spectra, the combination of the  energy output expected in hydrogen fusion  and  the  surface area of a typical WD giving the effective temperatures in the correct range, close to the observed values \citep{1992A&A...262...97V}.

Details of  thermonuclear burning  on the white dwarf surface are still a matter of debate.  Calculations by  \citet{2007ApJ...663.1269N,2013ApJ...777..136W} (and references therein)  find that hydrogen fusion  may proceed in a steady manner  above some value of the mass accretion rate, in the so called stability strip stretching around $\dot{M}\sim {\rm few}\times 10^{-7} ~M_{\sun}$ yr$^{-1}$,  its exact location depending on the WD mass. In this regime classical steady SSS are observed, while at lower mass accretion rates hydrogen fusion on the WD surface is unstable, giving rise to the explosions of Classical and recurrent novae.  On the other hand,  calculations by \citet{1995ApJ...445..789P,2005ApJ...623..398Y,2013IAUS..281..166S}, suggest that nuclear burning is unstable at any mass accretion rate.
Although the nova explosion itself is primarily  observed  in the optical band,  nuclear burning of the residual hydrogen on the WD surface after the explosion generates soft X-ray emission and the system is observed as a post-nova super-soft X-ray source  \citep{1995ASSL..205..453T,2013ApJ...777..136W,2016MNRAS.455..668S}. Thus, existence of super-soft X-ray sources, steady or not, is envisaged in either scenario.

Although accreting WDs are interesting objects on their own right, they became  of broader astrophysical importance as
possible progenitors of type Ia supernovae \citep[e.g.][]{1973ApJ...186.1007W,1982ApJ...253..798N}. However, before exploding as a supernovae, a white dwarf needs to increase its mass from the initial value, typically in the $\sim 0.5-1.0$ M$_\odot$ range, to near the Chandrasekhar mass limit of about $\sim 1.4$ M$_\odot$. For an accreting white dwarf, the only way to increase its mass is hydrogen and/or helium fusion on its surface. The mass growth is  most efficient when nuclear burning is stable,  as in all other regimes it is counteracted by various mass loss processes. However,   growth of the white dwarf mass is also possible in certain domains of the parameter space, when the burning is unstable \citep{2013IAUS..281..166S,2015MNRAS.446.1924H}.
Nevertheless, it was demonstrated that the number of super-soft X-ray sources \citep{2010ApJ...712..728D} and their total luminosity in the X-ray  \citep{2010Natur.463..924G,2011MNRAS.412..401B}  and UV \citep{woods13, johansson14, woods14} bands in nearby galaxies is by far insufficient to explain the observed SN Ia rates.

There is still no complete  understanding of how population of nuclear burning accreting white dwarfs is related to the fundamental properties of their host galaxies, such as their stellar mass, star-formation rate and star-formation history and metallicity.
On the theoretical side, there have been several attempts to address this problem by means of the population synthesis calculations  \citep[e.g.][]{1998ApJ...497..168Y,2010AstL...36..780Y}. Recently, \citet{2014MNRAS.445.1912C,2015MNRAS.453.3024C,2016MNRAS.458.2916C} implemented a hybrid population synthesis calculations using the MESA code \citep{2011ApJS..192....3P,2013ApJS..208....4P} to compute the mass transfer phase of the binary evolution. They have produced a model which was in a reasonable agreement with a number of existing constrains on the soft X-ray luminosities and average intensity of ionising radiation field, as well as with statistics of classical novae in nearby galaxies.

Observational studies of SSS populations are often impeded by the extreme softness of their spectra, making their search and characterisation difficult. Nevertheless,  there have been a number of successful attempts, which intensified with the advent of Chandra and XMM-Newton \citep[e.g.][]{2002ApJ...574..382S,2003ApJ...592..884D,2004ApJ...609..710D,2005A&A...442..879P,2010A&A...523A..89H,2011A&A...533A..52H,2014A&A...563A...2H}.
It was suggested that observed number of SSS on various types of galaxies are different \citep[e.g.][]{2010ApJ...712..728D}. \citet{2011MNRAS.412..401B} showed that specific frequencies of  SSS with $L_X>10^{36}$ erg/s in disks of late-type galaxies  by factor of $\sim$2 exceed that of bulges.

In this paper we continue the observational line of investigation of populations of accreting white dwarfs. To this end, we attempted to construct a suitable sample of galaxies from Chandra archive for search SSSs and implemented  efficient source classification algorithms.  Our goal is to identify steady nuclear burning white dwarfs and to   estimate their specific frequencies in early- and late- type galaxies.

The paper is structured as follows: In Section 2 we introduce  our sample of nearby galaxies observed by Chandra. Observations and data reduction procedures are described in Section 3. In Section 4 we introduce the initial source classification procedure based on combination of median energy and hardness ratio and in Section 5 we investigate  properties of soft X-ray sources and separate nuclear burning white dwarfs from the rest of source population with the help of   the temperature-luminosity diagram. The nature of soft and super-soft X-ray sources is described in Section 6 where we also estimate their specific frequencies and compare our results with previous work,  and in Section 7 we  summarise our findings.

\section{A sample of nearby galaxies}

\label{sec:sample_galaxies}

\begin{figure*}
\centering
\begin{tabular}{cc}
   \includegraphics[width=0.4\linewidth]{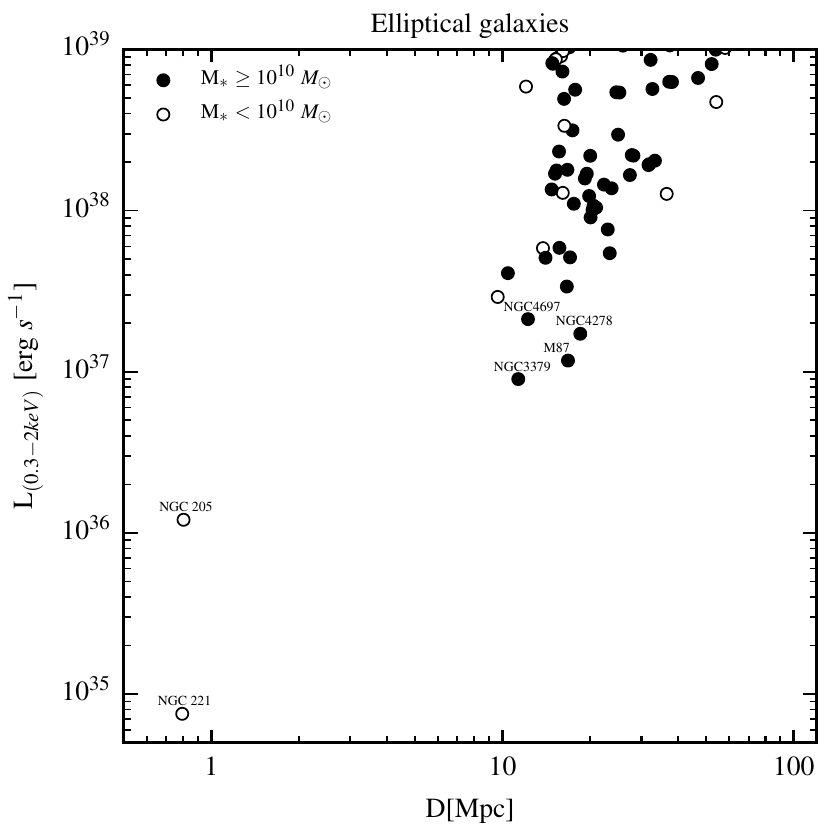}&
    \includegraphics[width=0.4\linewidth]{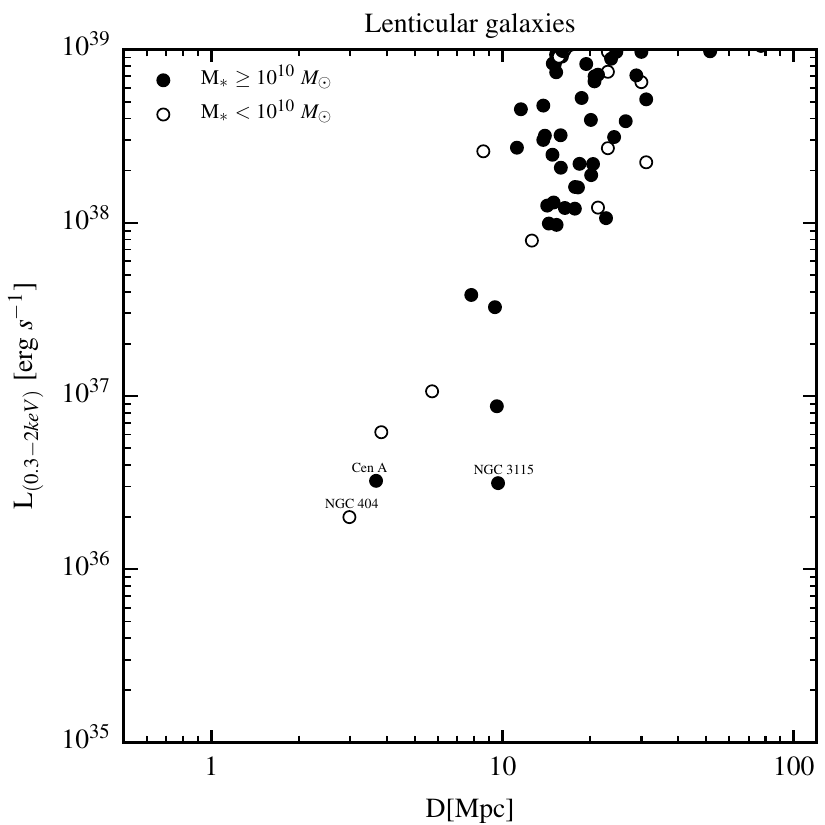}\\[2\tabcolsep]
    \includegraphics[width=0.4\textwidth]{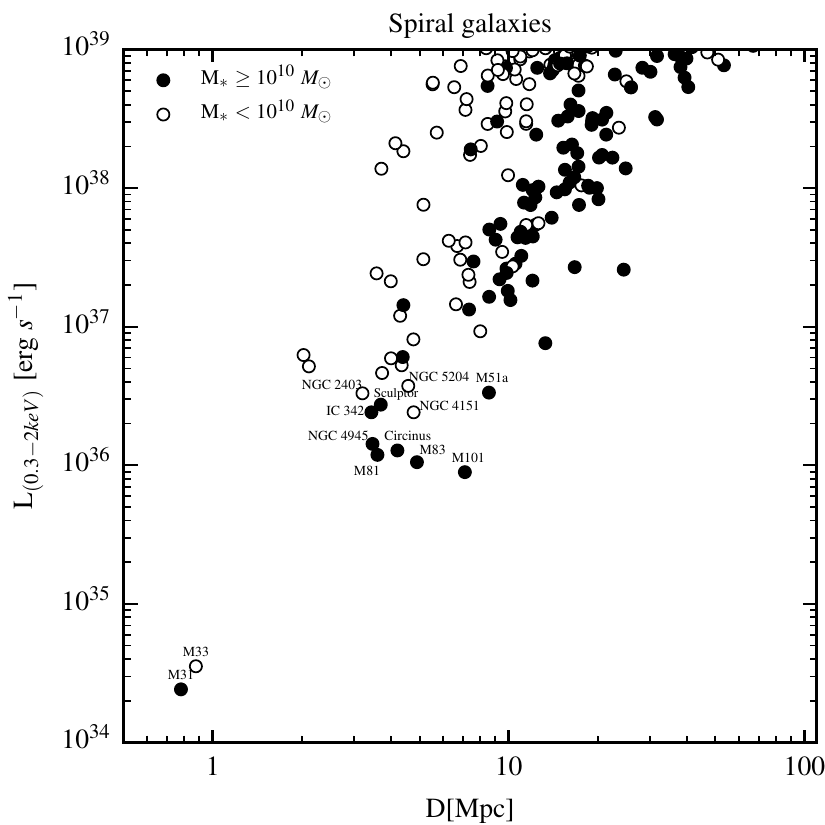} &
    \includegraphics[width=0.4\linewidth]{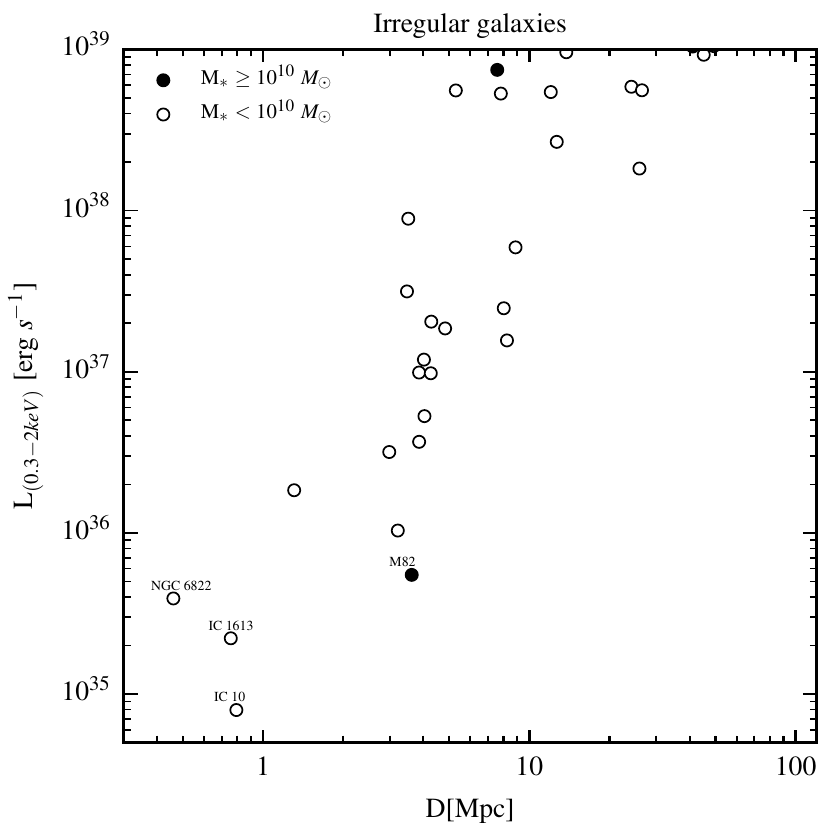}
 \end{tabular}
\caption{ The X-ray sensitivity--distance diagram of nearby galaxies observed by {\it Chandra} up to cycle 20. Each panel corresponds to a certain morphological type of galaxies. Filled  (open) symbols correspond  to galaxies with stellar mass $M_* \ge 10^{10}~M_\odot ~ (M_* <10^{10}~ M_\odot) $. See Section \ref{sec:sample_galaxies} for details.}
\label{fig:galaxies}
\end{figure*}

\begin{table*}
\caption{List of nearby galaxies used to study populations of soft and super-soft X-ray sources.
\label{tab:galaxy_sample}}
\renewcommand\arraystretch{1.5}
\centering
\begin{tabular}{lccccccc}
\hline
Name & Type & D         & Angular size                                 & (B--V)$_0$                                  & M$_*$                        & N$_{\rm H,Gal}$ & SFR\\
         &          & (Mpc)  & (D$_{25}$ $\times$ d$_{25}$; PA)  & RC3             &         ($\rm 10^{10}\ M_\odot$) & (10$^{20}$ cm$^{-2}$)  &  ($\rm M_\odot$/yr)\\
      (1)   &    (2)      & (3)  & (4)  & (5) & (6) & (7) & (8)\\
\hline
NGC 5236 (M83)                       &Sc        &4.6    &12.9$\arcmin$ $\times$ 11.5$\arcmin$; 54$\degr$        & 0.61    & 3.0   & 4.12 &3.1\\
NGC 5194 (M51)                       &SABb   &7.6    &11.2$\arcmin$ $\times$ 6.9$\arcmin$; 173$\degr$        &0.53     & 4.5   & 3.33 &2.9\\
NGC 5457 (M101)                     &SABc  &6.7  &28.8$\arcmin$ $\times$ 26.9$\arcmin$; 36$\degr$           &0.44    &2.3    &8.58  &2.9\\
NGC 3031 (M81)                       &Sab    &3.6   &26.9$\arcmin$ $\times$ 14.1$\arcmin$; 157$\degr$        &0.82     &5.4   &10.2  &0.5\\
NGC 5128 (Cen A)   			&S0      &3.4    &25.7$\arcmin$ $\times$ 20.0$\arcmin$; 35$\degr$        &0.88    &4.9    &2.35  &0.8\\
NGC 3115                                &E-S0   &9.7    &7.2$\arcmin$ $\times$ 2.5$\arcmin$; 40$\degr$            &0.94     &6.9   & 3.88  &--\\
NGC 3379               			&E        &10.6  &5.4$\arcmin$ $\times$ 4.8$\arcmin$; 71$\degr$            &0.94     &5.8  & 2.65 &--\\
NGC 4278                                &E         &16.1  &4.1$\arcmin$ $\times$ 3.8$\arcmin$; 22.7$\degr$        &0.90    &6.0  &2.04  &--\\
NGC 4697                                &E         &11.7  &7.2$\arcmin$ $\times$ 4.7$\arcmin$; 70$\degr$            &0.89    &6.0  &2.08 &-- \\
\hline
\end{tabular}
\flushleft
Notes: (1) -- Galaxy name.  (2) -- Morphological type from HyperLeda catalogue \citep{2014A&A...570A..13M}. (3) -- Distances to galaxies obtained by the following  methods: Cepheids:  M83 \citep{2006ApJS..165..108S}, M81 and M101 \citep{2001ApJ...553...47F}, Cen A \citep{2007ApJ...654..186F}; planetary nebulae luminosity function: M51 \citep{2002ApJ...577...31C}; surface brightness fluctuations: NGC 3379, NGC 3115, NGC 4278, NGC 4697 \citep{2001ApJ...546..681T}. (4) -- Major D$_{25}$,  minor d$_{25}$ diameters and position angles PA from RC3 catalogue \citep{1991rc3..book.....D}. For several galaxies PA were adopted from elsewhere: M83 \citep{2011ApJS..197...21H}; M51 \citep{2014ApJ...784....4C}; M101  \citep{2018ApJ...854...68H}; NGC 3379 and NGC 4278 \citep{1990AJ....100.1091P}. (5) -- The total colour indexes corrected for galactic and internal extinction from RC3 catalogue. (6) -- Stellar mass within D$_{25}$ of the galaxy obtained by mass-to-light ratio for K$_s$ band with colour index (B--V)$_0$ from RC3 using calibration from  \citet{2001ApJ...550..212B}. Total K$_s$  magnitudes were calculated using data from 2MASS Large Galaxy Atlas \citep{2003AJ....125..525J}.  Absolute K$_s$ magnitude of the Sun was assumed to be $K_{s,\odot}$=3.27 \citep{2018ApJS..236...47W}. (7) -- Galactic absorption column densities \citep{2016A&A...594A.116H}. (8) -- Star formation rates of spiral galaxies were adopted from \citet{2013AJ....145....6J}, for Centaurus A from \citet{2019ApJ...887...88E}. Values were corrected to the distance given in column (3).
\end{table*}

Galaxies for the analysis were selected from Chandra observatory archive for cycles 1--20.  To this end we selected observations from the following categories: normal galaxies, active galaxies and quasars, clusters of galaxies, galactic diffuse emission and surveys. Only archival observations with ACIS detector without grating were selected. For each galaxy, we grouped  observations made within same Chandra cycle  and selected the group having largest total exposure time. We then cross-matched aim-point positions of selected  {\it Chandra} observations with positions of galaxies in RC3 catalogue \citep{1991rc3..book.....D} to select potentially interesting observations. For each galaxy, we used the match radius of 0.5$\times$D$_{25}$. Parameters of galaxies required for the further selection were computed as follows. Stellar masses were computed from K$_s$ band luminosities using  mass-to-light ratios computed following the prescription of  \citet{2001ApJ...550..212B}. Classification of galaxies, distances and K$_s$-band magnitudes were taken from HyperLeda\footnote{http://leda.univ-lyon1.fr/} catalogue \citep{2014A&A...570A..13M} and Updated Nearby Galaxy Catalog \citep{2013AJ....145..101K}.  Absorption corrected (B-V)$_0$ colours were taken from RC3 catalogue.

\subsection{Selection criteria}
\label{sec:sample_galaxies_criteria}

We built our selection procedure based on  the following considerations.

{\it i) Stellar mass of the host galaxy.}
Population synthesis calculations of \citet{2014MNRAS.445.1912C,2015MNRAS.453.3024C}  predict specific frequencies of observable (apparent luminosity $L_X>10^{36}$ erg/s) stable burning accreting WDs in spiral galaxies after  10 Gyrs of evolution   at the level of $\sim 4\times 10^{-10}$ ($\sim 2\times 10^{-9}$) sources per $M_\odot$, assuming absorbing column density of $3\times 10^{21}$ ($3\times10^{20}$) cm$^{-2}$. For  elliptical galaxies their prediction is $\sim 10^{-10}$ sources per $M_\odot$ for hydrogen column density $3\times10^{20}$ cm$^{-2}$.
Based on these predictions we will be only considering sufficiently massive galaxies with stellar masses $M_{*}$ $\ga$ $10^{10}$ $M_\odot$, in  which we can  expect to find $\ga 1$ nuclear burning accreting WD.

{\it ii) Distance.}  We excluded from our analysis a few large nearby galaxies which  do not fit, by a large margin, in the Chandra field of view and require analysing mosaics of a large number of  observations. These  are: Magellanic Clouds, M31 and M33. We should also exclude too distant galaxies which will suffer from confusion. However, they are automatically excluded based on the sensitivity arguments (item iv below)

{\it iii) Inclination angle.}
Source populations in spiral galaxies  which are viewed close to edge-on will be obscured by  large intrinsic absorption by gas and dust in the disk, essentially reducing the numbers of expected super-soft sources to zero. Therefore  in selecting late-type galaxies we applied a criterion $i<70$ degrees. For this reason,  for example,  Sculptor and NGC 4945 were not included in our final sample. No inclination cut was applied to early-type galaxies.

{\it iv) Sensitivity limit of available archival data.}
The  sensitivity limit of  archival data can be estimated as follows:
\begin{eqnarray}
L_X\approx 4\pi D^2\times ECF \times \frac{S}{t},
\label{eq:Lx_lim}
\end{eqnarray}
where  {\it D} is distance to the galaxy, $ECF$ is counts-to-ergs  conversion factor, $S$ is the limiting number of source counts required to detect and characterise a source, {\it $t$} is total exposure time of archival data. We assumed black body model with temperature 60 eV and require $S=20$ counts to detect and classify a source.
For each galaxy we obtained $ECF$ from black body model ($kT=60$ eV) modified by  the Galactic absorption column density in the direction of the  galaxy. For the latter we used the HI4PI  data \citep{2016A&A...594A.116H}. For cycles 7--20, in computing  ECF, we used the ACIS detector aim-point response files from {\it Chandra} Calibration Database\footnote{ftp://cxc.cfa.harvard.edu/pub/caldb/ }. For cycles 1--6  we generated aim-point response files ourselves. Circinus ($5.3\times10^{21}$ cm$^{-2}$)  and IC 342 ($3.6\times10^{21}$ cm$^{-2}$) galaxies with high Galactic absorption column densities  were excluded from our sample. We also excluded  M87 and M82 due to the presence of bright and complex diffuse emission  in these galaxies.

We plot the distribution of galaxies of various morphological types on the distance -- sensitivity limit plane in Fig.\ref{fig:galaxies}. Based on these data we selected spiral galaxies M51, M81, M83, M101, lenticular galaxies Cen A and NGC3115 and elliptical galaxies NGC 3379, 4278 and 4697. For  elliptical galaxies, sensitivity limits of the archival Chandra data exceed $10^{37}$ erg/s, this is however compensated by zero intrinsic absorption in these galaxies, see Section \ref{sec:temp}.

We experimented with the selection criteria in order to check if our sample can be improved by adding a few  galaxies lying on the boundary of our luminosity and stellar mass selection. We found, however, that  relaxing the sensitivity threshold we need to raise the luminosity cut on the $kT_{bb}-L_X$ plane (Section \ref{sec:temp}) which actually leads to the loss of super-soft sources of lower luminosity and degradation of the statistical accuracy of our analysis.  Adding less massive galaxies, below the assumed cut of $10^{10}$ $M_\odot$, on the other hand, does not lead to the quick improvement of the sensitivity to the populations of SSS as the total mass already included in the analysis exceeds $\approx 4\times 10^{11}$ $M_\odot$.  We therefore conclude that the list of galaxies assembled here is reasonably optimal for the current analysis.
Basic parameters of these galaxies are summarised in Table \ref{tab:galaxy_sample}.

\section{Observations and Data Reduction}
\label{sec:obs_dat}

The observations were processed by following a standard CIAO \citep{2006SPIE.6270E..1VF}  threads\footnote{http://cxc.harvard.edu/ciao/index.html} (CIAO version 4.8 and CALDB version 4.7.1). We used tool {\it chandra\_repro} to reprocess data and create level 2 event file. To improve absolute astrometry  we used {\it wcs\_match} and {\it wcs\_update}. Astrometry correction was done relative to the observation with the longest exposure time using positions  of bright X-ray sources to match frames. Event files have been merged by {\it reproject\_obs} with aim point position of observation with the longest exposure taken as the reference. Combined images and  exposure maps were obtained running {\it flux\_obs}. In Table \ref{tab:short_obs} we present short information about observations used in analysis. Detail information available in Appendix \ref{appendix:long_obs}.

\begin{table*}
\caption{Properties of galaxies and statistics of their X--ray source populations.
\label{tab:short_obs}}
\renewcommand\arraystretch{1.5}
\centering
\begin{tabular}{lcccccccccc}
\hline
Name                   & Date                                     & Cycle  & \#   &Exposure & L$_{\rm 0.3-2}^{\rm faint }$          & N$_{\rm src}$   & N$_{\rm src}^{\rm soft}$ &  N$_{\rm src}^{\rm SSS}$  & M$_*$               &  \\
                            &                                             &         &          & (ksec)     & (10$^{35}$ erg/s) & &  &       & $(10^{10} M_\odot$) \\
                     (1)   &                                      (2)   & (3)      & (4)   & (5)         & (6)                       & (7)              & (8)              &(9)                    & (10) \\
\hline
NGC 5236 (M83)   &23/12/2010 - 28/12/2011  & 12      & 10   &729.63    &3.0     &276          &116               &14     &2.5      \\
NGC 5194 (M51)   &09/09/2012 - 10/10/2012  &13       & 7     &745.33    &5.8     &245          &105              &5       &4.1     \\
NGC 5457 (M101) &19/01/2004 - 01/01/2005  &5         &24    &952.01    &6.3     &221           &89               &11     &2.0     \\
NGC 3031 (M81)   &26/05/2005 - 06/07/2005  &6         &15    &175.92    &3.5      &161    	    &50               &2       &5.3     \\
NGC 5128 (Cen A) &22/03/2007 - 30/05/2007  &8         & 6     &561.72   &11.7    &286          &88    		&2       &3.6     \\
NGC 3379   &23/01/2006 - 10/01/2007  &7         & 4     &305.53    &20.1    &87            &30     		& 1      &5.8     \\
NGC 3115             &18/01/2012 - 06/04/2012  &13       & 8     &985.40    &7.7       &136          &32     		& 0      &6.9    \\
NGC 4278             &16/03/2006 - 20/04/2007  &7         &5      &433.30    &36.0     &244          &92      	&2       &6.0      \\
NGC 4697             &26/12/2003 - 18/08/2004  &5         &4      &153.70    &34.7     &102          &38       	&2       &6.0       \\
\hline
\end{tabular}
\flushleft
Notes: (1) -- Galaxy name. (2) -- Time interval of observations. (3) --  {\it Chandra} cycle during which most of observations were made. (4) and (5) -- Total number of observations and their exposure time. Full information about archival data used in this work are available in Appendix \ref{appendix:long_obs}. (6) -- X-ray luminosity of faintest source detected in combined data. In computing this luminosity absorbed black body model was used  with temperature $kT=60$ eV and absorption column density computed as described in  Section \ref{sec:nh}. (7) -- Total number of detected sources  (see Section  \ref{sec:src_det}). (8) -- Total number of  sources classified as soft according to the method described in Section 4, and having $\ge$ 20 net counts. (9) -- Number of  super-soft X-ray sources -- sources located to the left of the stable hydrogen burning boundary in Fig.\ref{fig:lum_kt} (see Section  \ref{sec:SSS_sample} and Appendix \ref{table:catalog}). (10) -- Stellar mass within the spatial region used for source detection.

\end{table*}
\label{appendix:image}
\label{appendix:long_obs}

\subsection{Source detection}
\label{sec:src_det}

Source detection was performed with {\it wavdetect} tool. We chose to use small $\sqrt{2}$ scale series with power of 0--3 of the wavelet functions to minimize the spurious detections associated with soft diffuse emission in galaxies. Scales of wavelets correspond to variation of the point spread function (PSF) within an aim point of single ACIS chip. We set up following parameters of {\it wavdetect} to create a background map from merged image: {\it bkgsigthresh}=0.01, {\it maxiter} = 10, {\it iterstop} = 10$^{-6}$. The false source detection probability threshold was chosen at  $10^{-7}$. Mask filtering were applied to all images of galaxies using exposure maps. Pixels were included to the  source detection region where exposure map values were greater than 20\% of the maximum ({\it expthresh} =0.2). False colour X-ray images of all galaxies with their D$_{25}$ and source detection regions are presented in Appendix \ref{appendix:image}.

To maximise sensitivity of source detection, we used combination of the following  energy bands: 0.3--1 keV; 1--2 keV; 0.3--2 keV; 0.5--8 keV; 2--8 keV bands. We ran {\em wavdetect} in each of these bands and merged their results into a single source list, excluding duplications. To this end we used  STILTS package \citep{2006ASPC..351..666T}, with the match radius equal to the  sum of their PSFs radii defined at the 80\%  encircled  fraction. In the final merged list we chose coordinates of sources in the band where the significance of detection was maximal.  In the very inner regions of some galaxies {\em wavdetect} found multiple overlapping sources. Visual inspection of the images showed that many of these sources are likely false detections caused by presence of sub-structures in the bright and complex  diffuse emission of the nucleus. In many cases it appears impossible to separate true compact  sources from  such false detections. For this reason we decided to exclude in M51, M83 and Centaurus A  galaxies their central regions of radius of  $15\arcsec$. In the case of Centaurus A we also excluded regions around bright X-ray jets similar to \citet{2006A&A...447...71V}.

To define source and background counts extraction regions we produced  PSF maps. In order to account for PSF variations across the combined image, we created  PSF map for each observation with {\it mkpsfmap} tool using encircled energy of 80\%, and combined individual maps weighting them with respective exposure maps. Combined source spectra used in the further analysis were extracted using CIAO tool {\it specextract}. For  source regions we used circles with the radius equal to the 80\% PSFs radius from the exposure-weighted PSF map. The background regions were defined as annuli with the inner and outer radii equal to two and four times the PSFs radius respectively. The source count rates were corrected for the source counts leakage to the background region. When source or background regions of different sources overlapped, we excluded overlapping parts from the counts summation, correcting the PSF fractions accordingly.

The lists of  detected sources were analysed to identify spurious detection due to fluctuations of local background. For each sources we tested the null hypothesis that observed counts in source and background regions are Poissonian realizations  of local background with the same mean \citep[][see Appendix A]{2007ApJ...657.1026W}. We estimated probabilities using source and background counts from merged images in 0.3--2 keV band.  All detections of statistical confidence of less than  3$\sigma$ were excluded from the final source list.  Additionally,  an  eye inspection of  X-ray images has not revealed any potential spurious detections. In Table \ref{tab:short_obs} we summarise numbers of X-ray sources detected in galaxies of our sample.

\subsection{Stellar masses of the final sample of galaxies}

Stellar masses of galaxies  within D$_{25}$ and within the source detection regions listed in Tables \ref{tab:galaxy_sample} and \ref{tab:short_obs} were determined following prescription of \citet{2012A&A...546A..36Z} as summarised below. We used  background subtracted images from 2MASS Large Galaxy Atlas \citep{2003AJ....125..525J}. Bright sources were visually removed from the images. Net counts of the images S were converted into magnitude using zero-point magnitude KMAGZP from a headers of fits files.
\begin{equation}
m_{\rm K} ={\rm KMAGZP}-2.5\ {\rm log(S)}
\end{equation}
We obtained K$_s$ band absolute magnitudes  using distances listed in Table \ref{tab:galaxy_sample}. In computing luminosities,  Solar absolute magnitude in K$_s$ band was assumed to be $K_{s,\odot}$=3.27 \citep{2018ApJS..236...47W}. Final stellar masses of the galaxies were obtained using K$_s$ band mass-to-light ratio \citep{2001ApJ...550..212B}.
\begin{equation}
{\rm log}(M_{\rm *}/L_{\rm K})=-0.692+0.652(B-V)_{\rm 0},
\end{equation}
where $L_{\rm K}$ is K$_s$ band luminosity and (B-V)$_{\rm 0}$ is colour index corrected for galactic and internal extinction (see Table \ref{tab:galaxy_sample}).

\begin{figure*}
\begin{center}
{
\includegraphics[width=58mm]{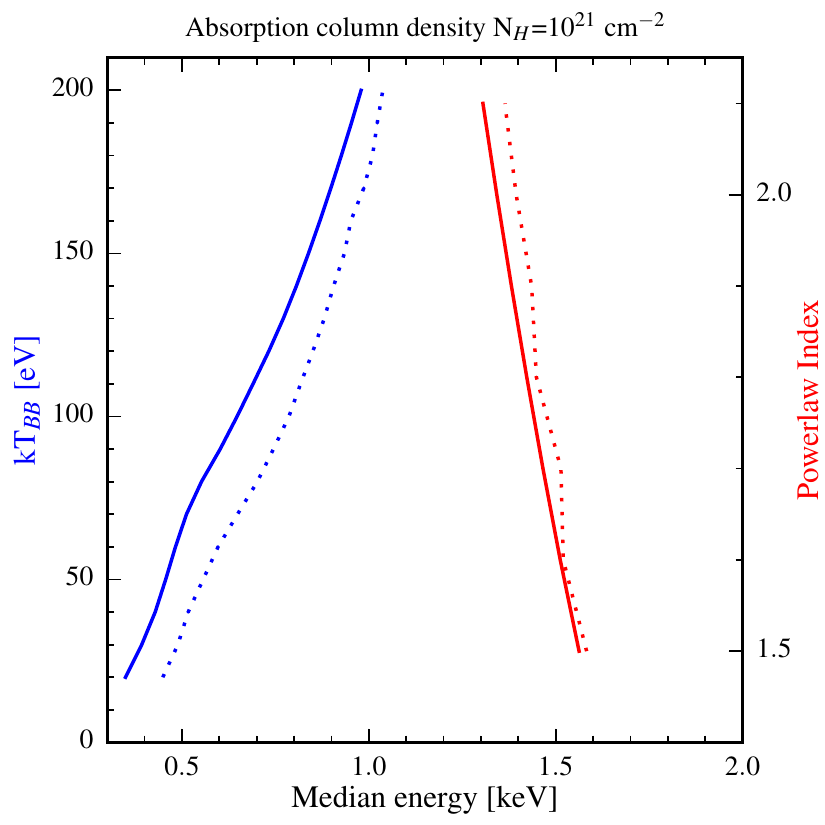}
\includegraphics[width=58mm]{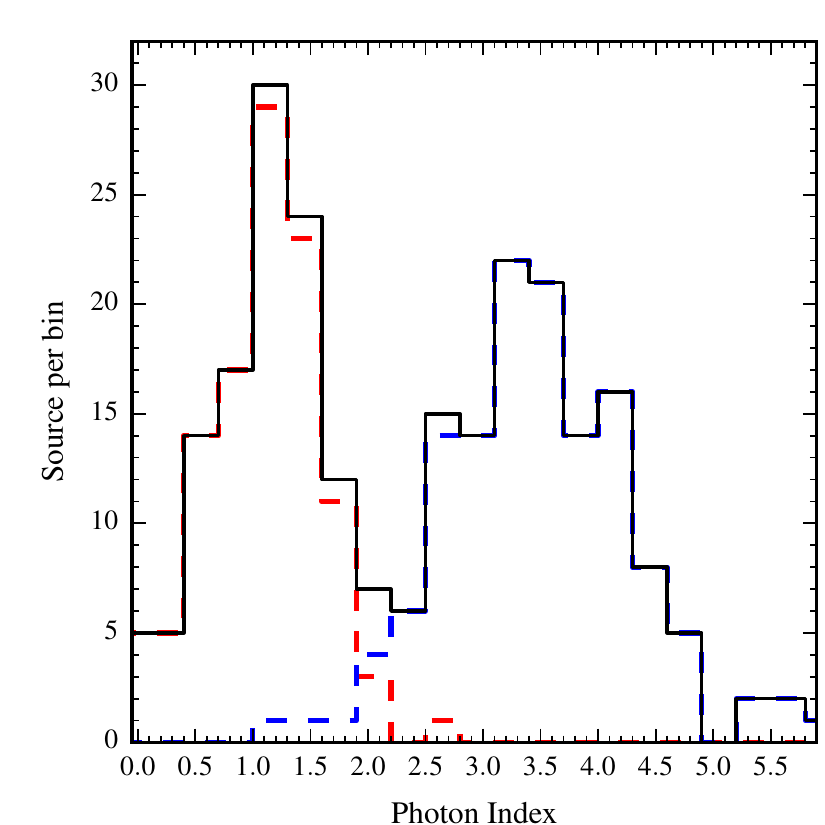}
}
\caption{{\it Left}: Relation of the median energy to the temperature of a black body model (left-hand curves and  y-xis, blue in the color version of this plot) and to the photon index of a power law model (right curves and y-axis, red). Simulations for Chandra ACIS-S detector with Cycle 13 (2012) response. Absorption column density of $10^{21}$ cm$^{-2}$ was assumed. {\it Solid lines} show to models  without background, {\it dotted lines}  --   models with 50\% contribution of  background counts.  Parameter ranges are relevant to super-soft sources and X-ray binaries, and the two types of spectra are clearly separated at about $\tilde{E} \approx 1.0-1.3$ keV.
{\it Right}: Verification of the source classification procedure in M51. Distribution of sources having median energy $\tilde{E}\le 2$ keV over photon index of power law model. {\it The blue and red dashed} histograms show distributions of sources classified as soft and hard, the {\it black solid line} is the total distribution. Bi-modality of the distribution of sources over photon index is obvious.}
\label{fig:model_median}
\end{center}
\end{figure*}

\section{Source classification}

\subsection{The method}
\label{hardness_median}

To classify and to separate soft and super-soft  sources from  sources with more usual spectra we used combination of hardness ratio $HR$ and median energy $\tilde{E}$. The hardness ratio was defined as a ratio of the  hard band count rate to the total broad band count rate:
\begin{eqnarray}
HR=\frac{R(2-8~keV)}{R(0.3-8~keV)}\ ,
\label{eq:hardness_ratio}
\end{eqnarray}
where $R$ is the background subtracted count rate in the respective energy band. Statistical error of the hardness ratio  was calculated using standard error propagation.

Median energy $\tilde{E}$ for each source was calculated from its observed counts spectrum considering the 0.3--8.0 keV energy band. The statistical error was  computed using the standard  formula for the variance of the sample median \citep{1960JASA..55..148R} where the probability density  at the median energy was approximately determined from observed counts distribution. In computing the median energy we chose not to subtract background from the source spectrum. Simulations have demonstrated that for the typical source and background spectra we are dealing with, this does not affect our source classification procedure (see Fig. \ref{fig:model_median} and its discussion below).

The motivation behind our choice  of the  parameters used for  source classification is fairly obvious. The median energy characterizes overall softness of the spectrum. The Fig.\ref{fig:model_median} (left panel) shows median energies of absorbed black body and power law models. Temperatures and photon index axes are located on the right and on the left side of Fig.\ref{fig:model_median} (left panel). As one can see  from the plot, in the parameter range relevant to nuclear-burning accreting white dwarfs ($kT\la 200$ eV) and X-ray binaries/AGN/stars ($\Gamma \la 2$), the two types of sources occupy different domains along the  $\tilde{E}$ axis with some gap  in between\footnote{Note that particular location of the boundary between super-soft sources and X-ray binaries in Fig.\ref{fig:model_median} depends on the hydrogen column density $N_H$ and details of the instrument response. The curves plotted in Fig.\ref{fig:model_median} (left panel) have been computed for 2012 (Cycle 13) Chandra ACIS-S response. For other Chandra calibration epochs their shape and location along x-axis will be (slightly) different.}.
Therefore one may expect a bi-modal distribution of sources over $\tilde{E}$ (which may however be smeared by statistical uncertainties).

We used M51 to illustrate and verify  classification of sources using their median energy (see corresponding panels  in Fig.\ref{fig:HR_MED}). To this end we fit background subtracted spectra of all sources  having $\tilde{E}\le 2$ keV\footnote{Some of the hard sources are located  at $\tilde{E}\ga 2$ keV, too large for a power law spectrum with slope of $\Gamma\la 2$  (Fig.\ref{fig:model_median}, left panel).
These are either sources with genuinely large $\tilde{E}$, e.g. absorbed AGN, or faint sources where contribution of the instrumental background  becomes important and shifts the median energy to non-physically  large values.}, with an absorbed power law model. The absorption column density for each source was fixed as described in Section \ref{sec:nh} below.  Distributions of soft and hard X-ray sources over photon index are shown in the right panel of  Fig.\ref{fig:model_median}. From these  distributions it is clear, that indeed, the sources classified as soft and hard occupy  different domains along the photon index axis. Small overlap of the two  distributions  is caused by statistical uncertainties in measuring the median energy and photon index.

This comparison shows that a classification scheme based on the median energy  is efficient in identifying  sources with soft spectra.
The hardness ratio, on the other hand, characterizes the presence  of  source emission above 2 keV and helps to identify  sources  having  low $\tilde{E}$ but also featuring hard emission components   extending to higher energies. Such hard spectral components are not typical for nuclear burning white dwarfs and may be encountered, for example,  in accreting black holes in the soft state and in some of supernova remnants.

Finally, we note that  the median energy is not efficient in separating the bulk of SNRs from white dwarfs. This can be achieved with more detailed spectral fitting, which however produces meaningful results only for the brightest sources.

Use of the median energy and hardness ratio for characterization of the spectral shape is not entirely new in X-ray astronomy, for example \citet{2004ApJ...614..508H} proposed to use various quantiles and \citet{2003ApJ...595..719P} used color-color diagram for this purposes. \citet{2003ApJ...592..884D} used hardness ratio to classify super-soft X-ray sources in external galaxies. Color-color diagrams and hardness ratio techniques were used in numerous other work not mentioned here. However, to our knowledge, combination of the median energy and hardness ratio was employed for separating super-soft sources from the bulk of the population for the first time in this paper.

\subsection{\boldmath Distribution of sources in $HR-\tilde{E}$ plane}
\label{sec:emed-hr}

\begin{figure*}
\begin{center}
\hbox
{
\includegraphics[width=58mm]{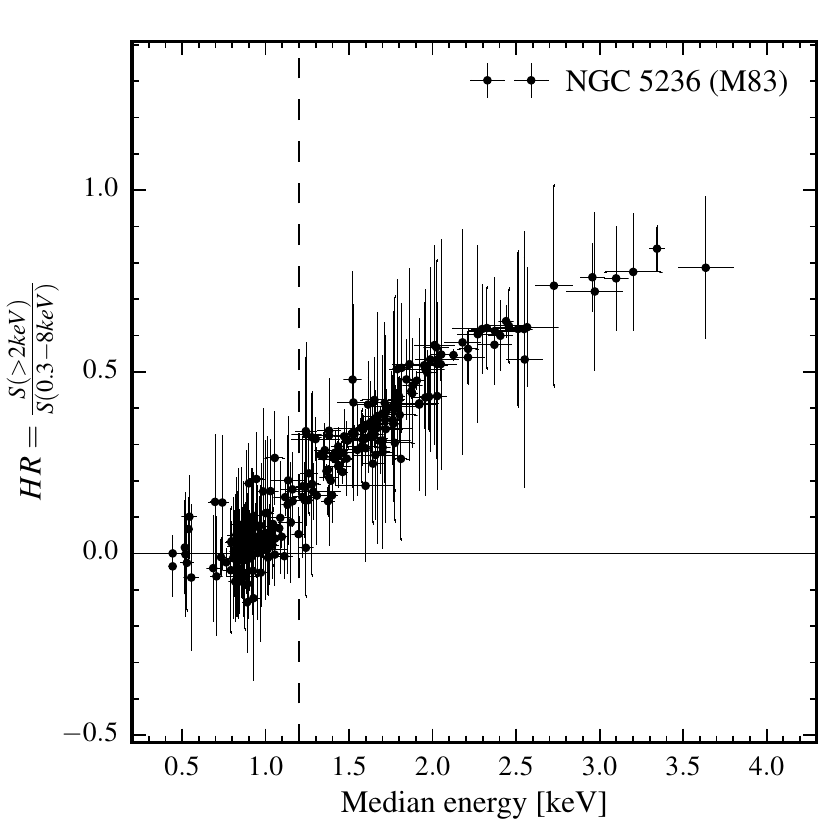}
\includegraphics[width=58mm]{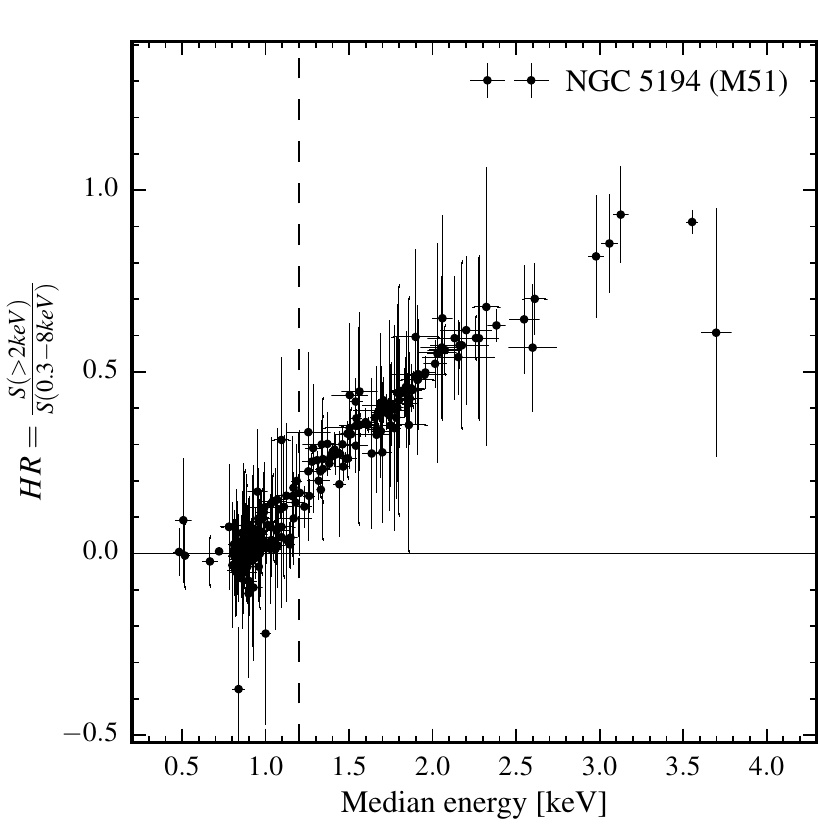}
\includegraphics[width=58mm]{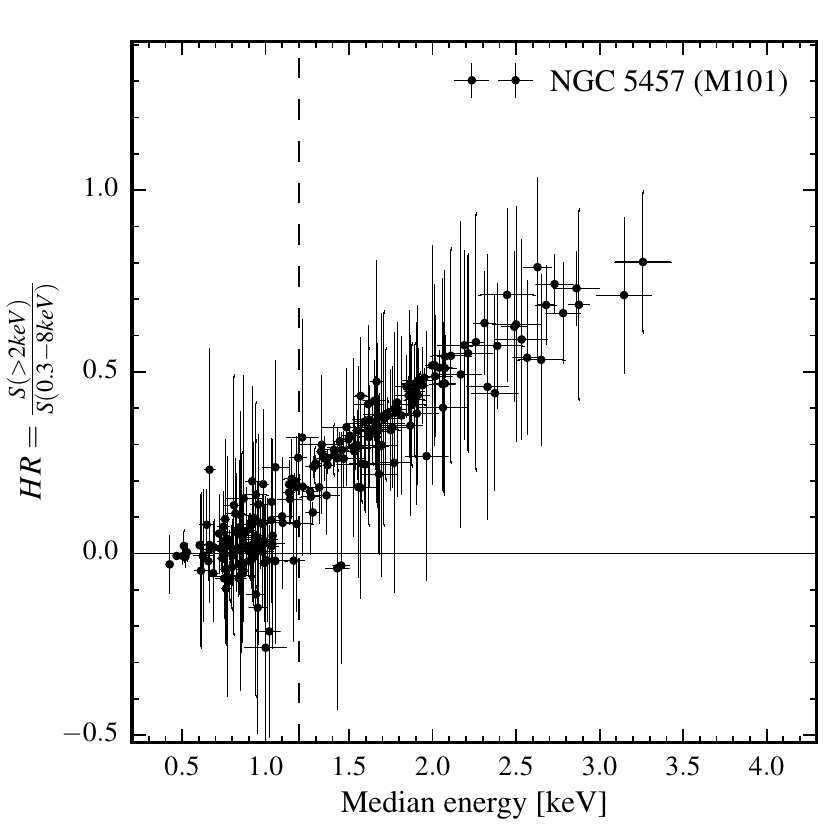}
}
\hbox
{
\includegraphics[width=58mm]{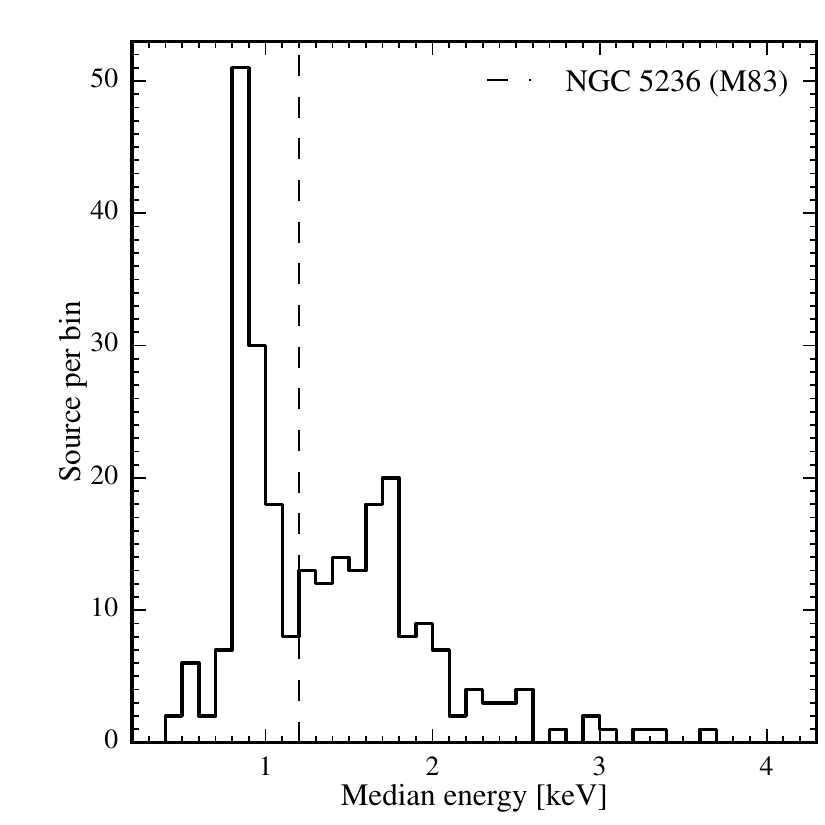}
\includegraphics[width=58mm]{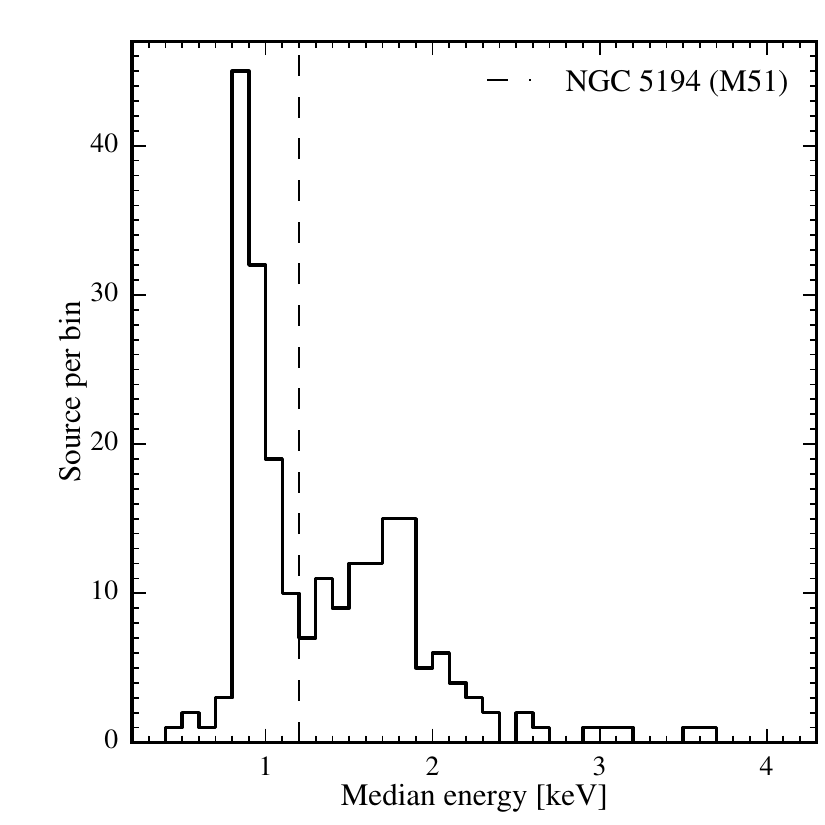}
\includegraphics[width=58mm]{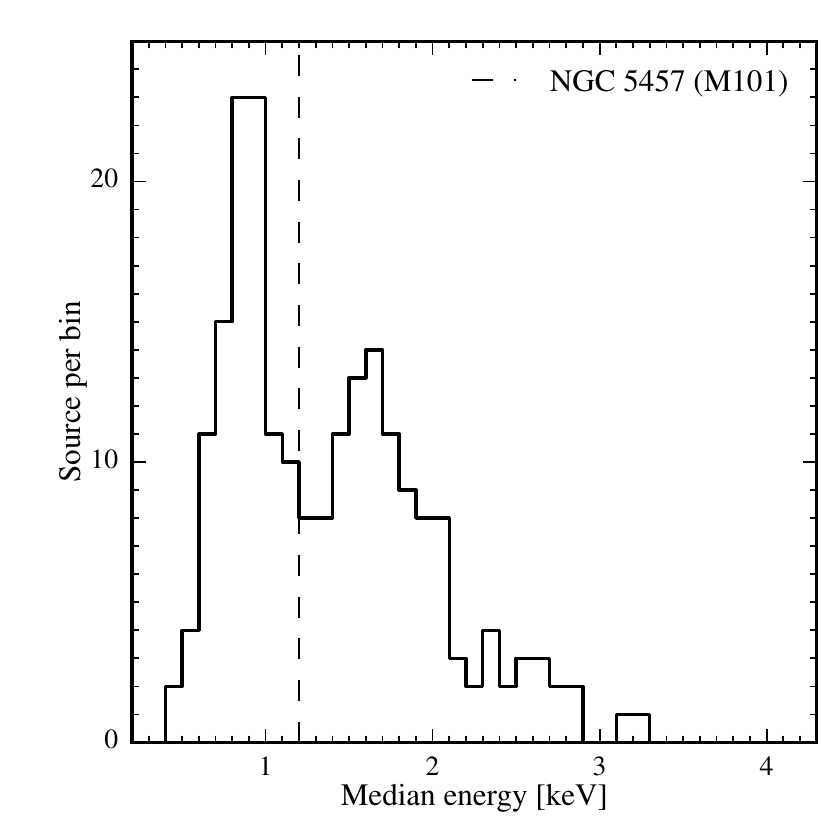}
}
\hbox
{
\includegraphics[width=58mm]{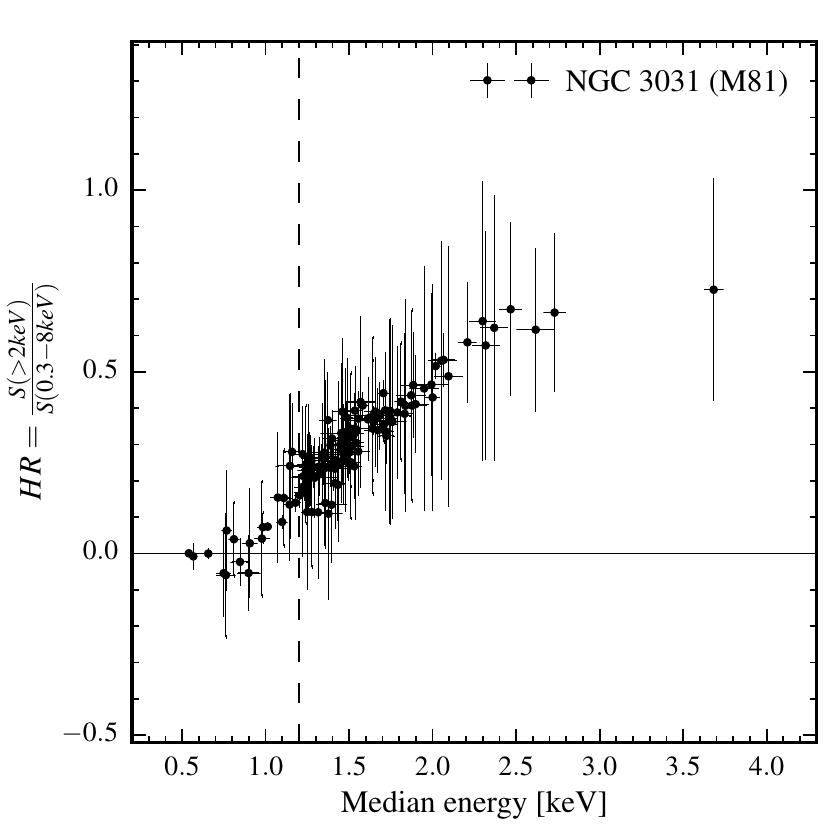}
\includegraphics[width=58mm]{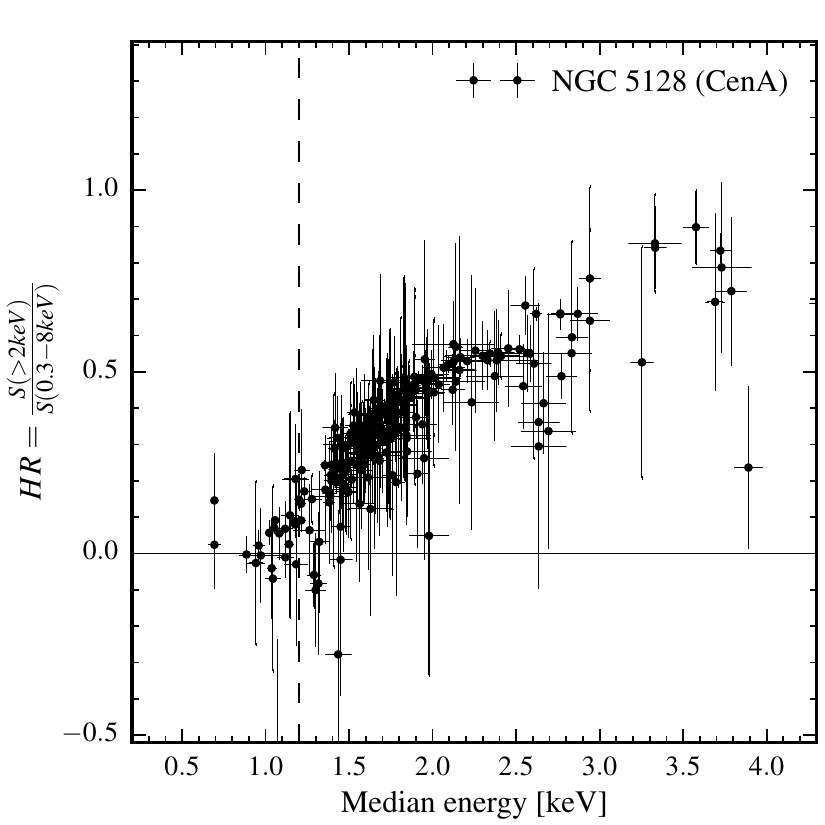}
\includegraphics[width=58mm]{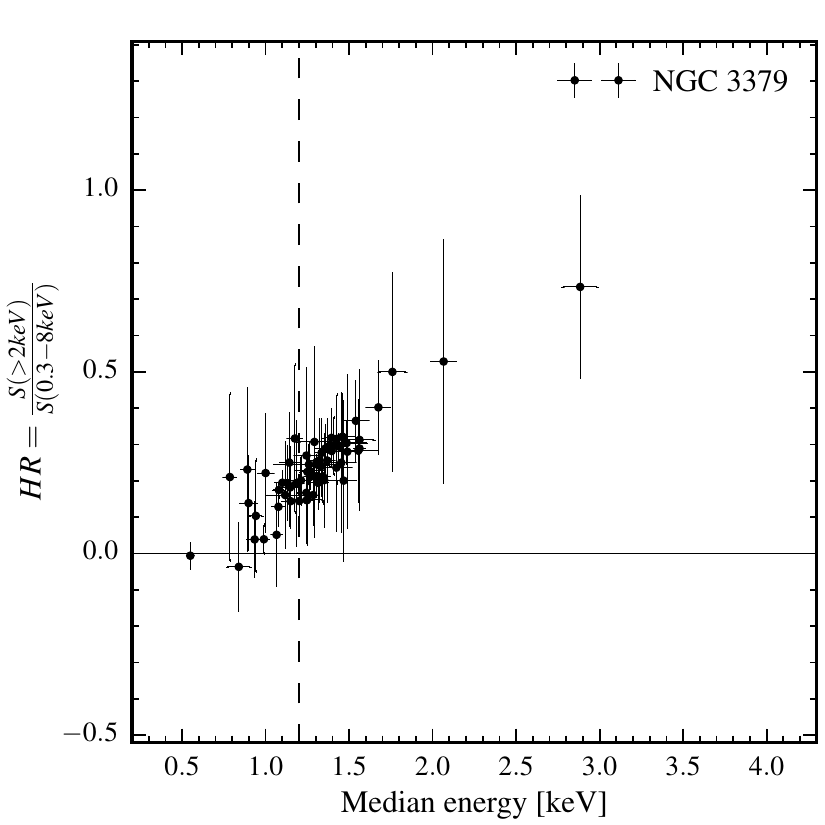}
}
\hbox
{
\includegraphics[width=58mm]{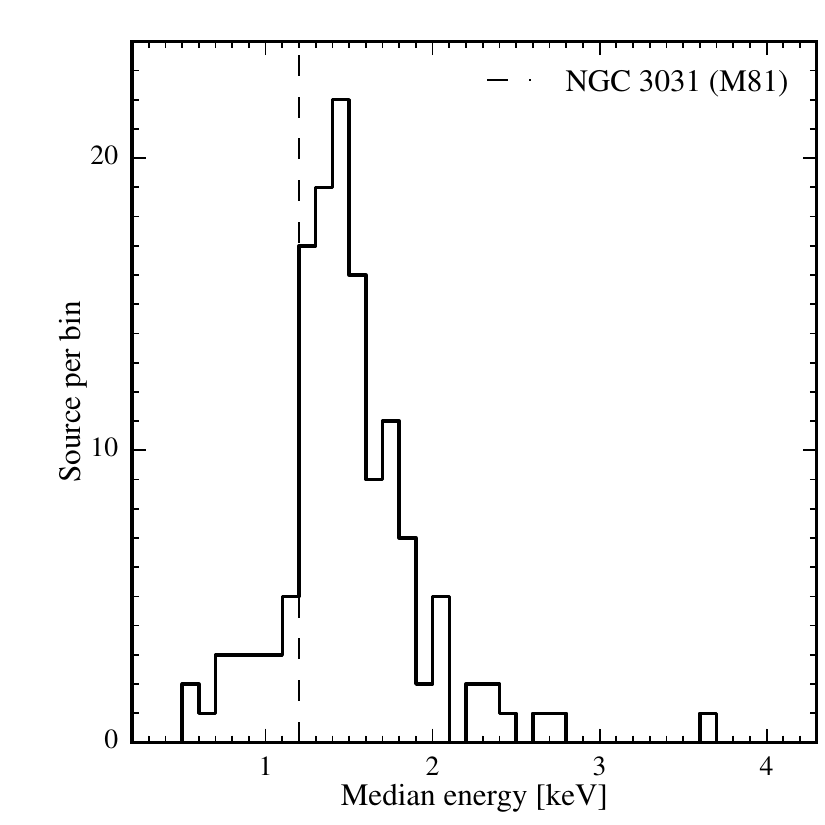}
\includegraphics[width=58mm]{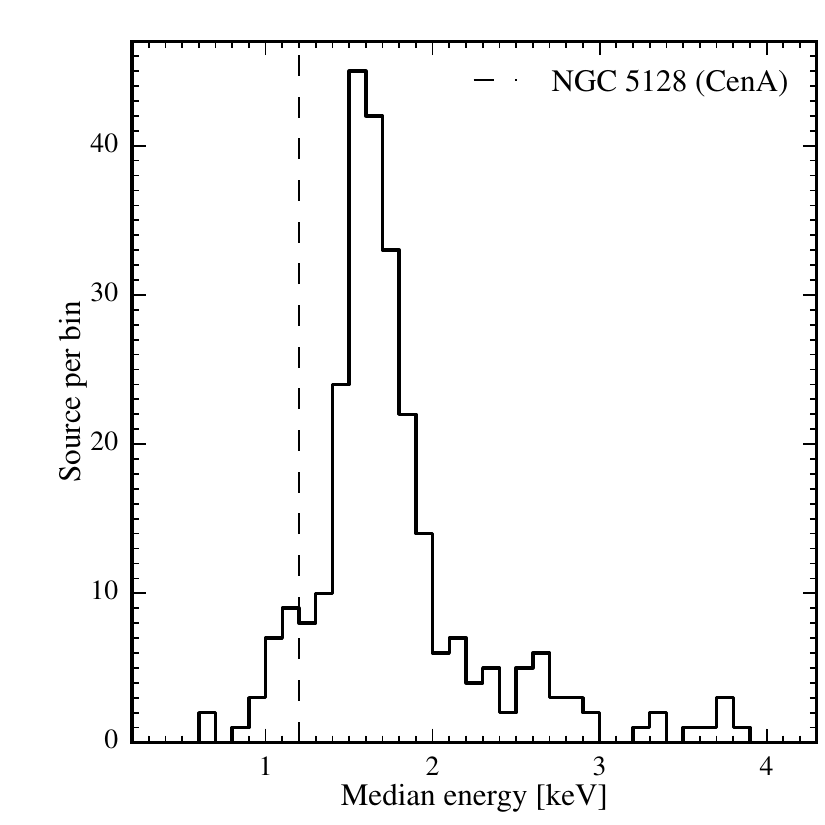}
\includegraphics[width=58mm]{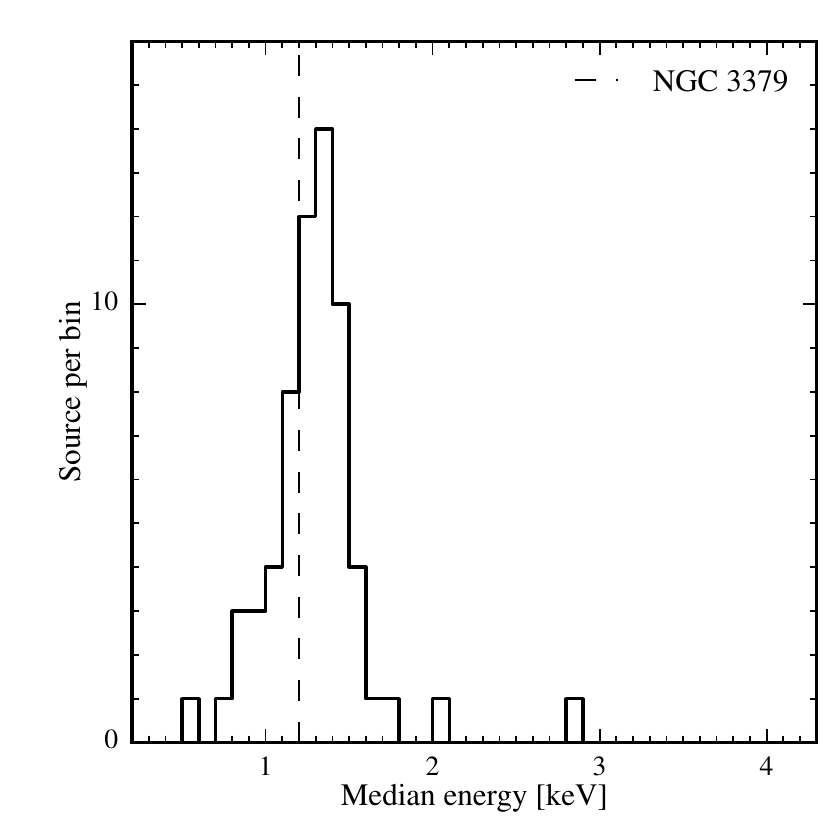}
}
\caption{{\it -- continued}}

\end{center}
\end{figure*}

\addtocounter{figure}{-1}

\begin{figure*}
\begin{center}
\hbox
{
\includegraphics[width=58mm]{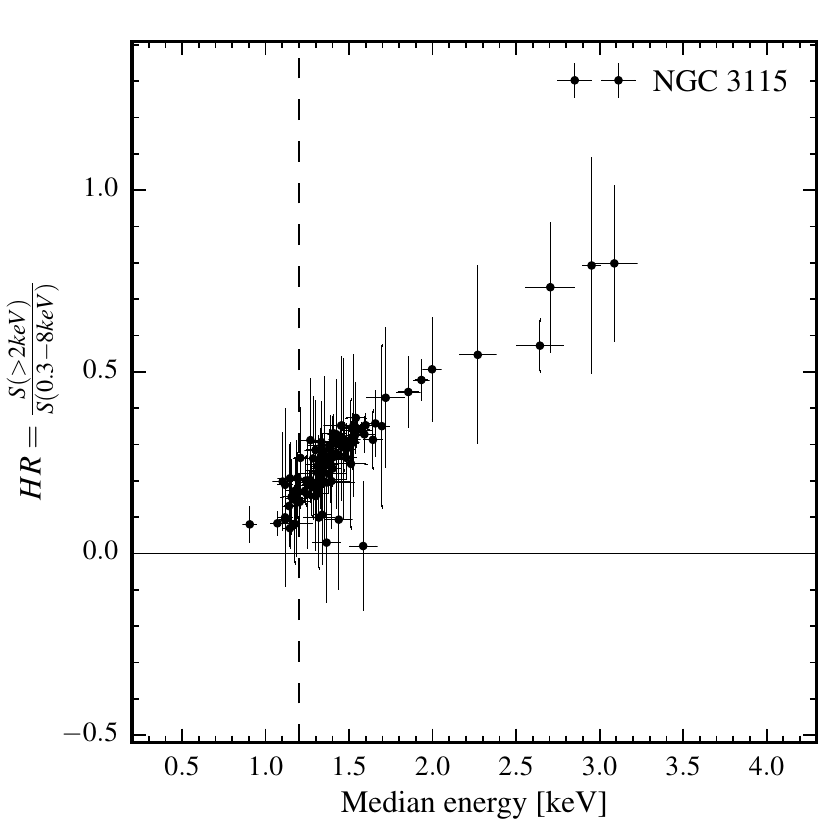}
\includegraphics[width=58mm]{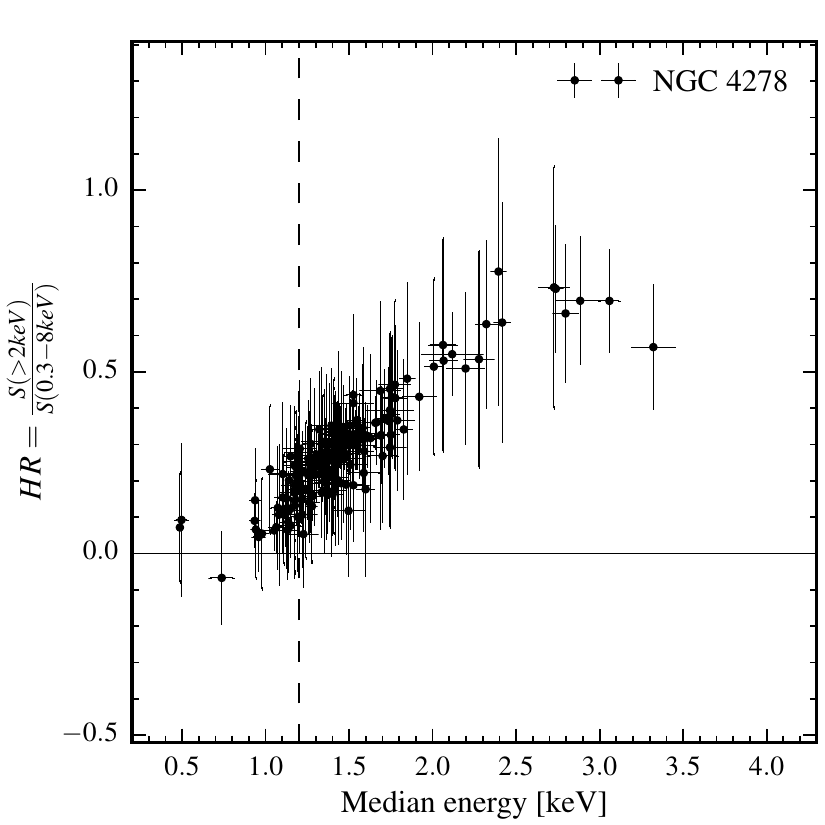}
\includegraphics[width=58mm]{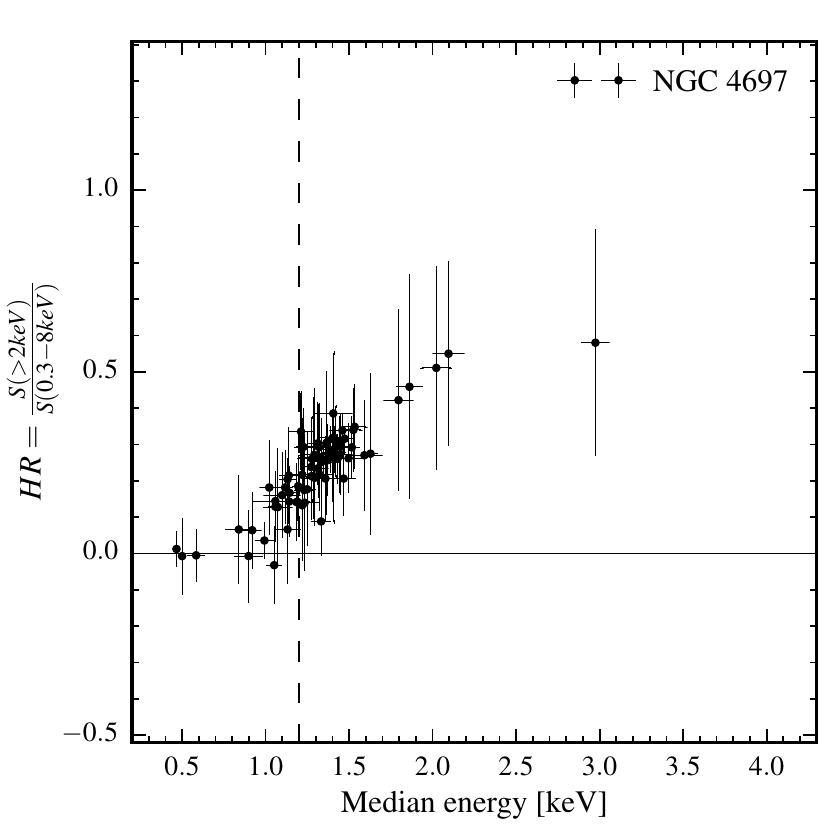}
}
\hbox
{
\includegraphics[width=58mm]{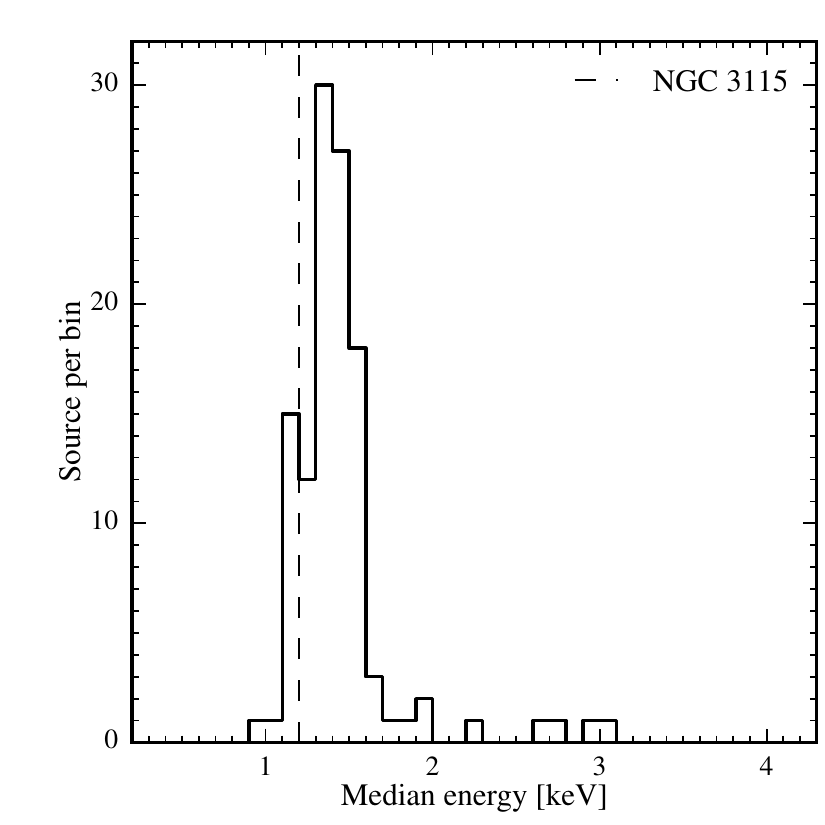}
\includegraphics[width=58mm]{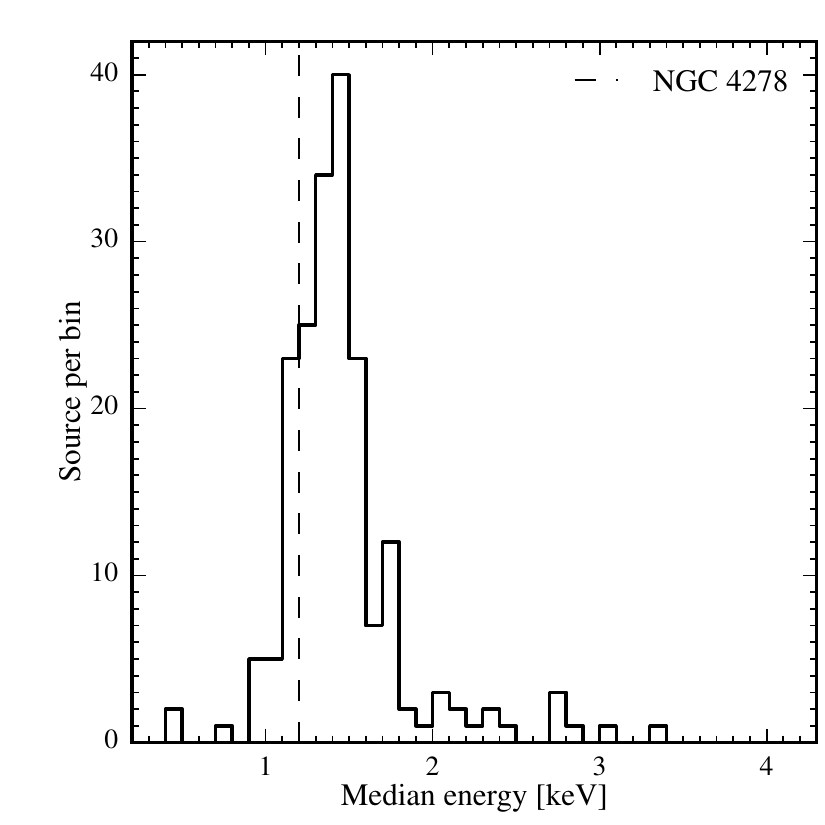}
\includegraphics[width=58mm]{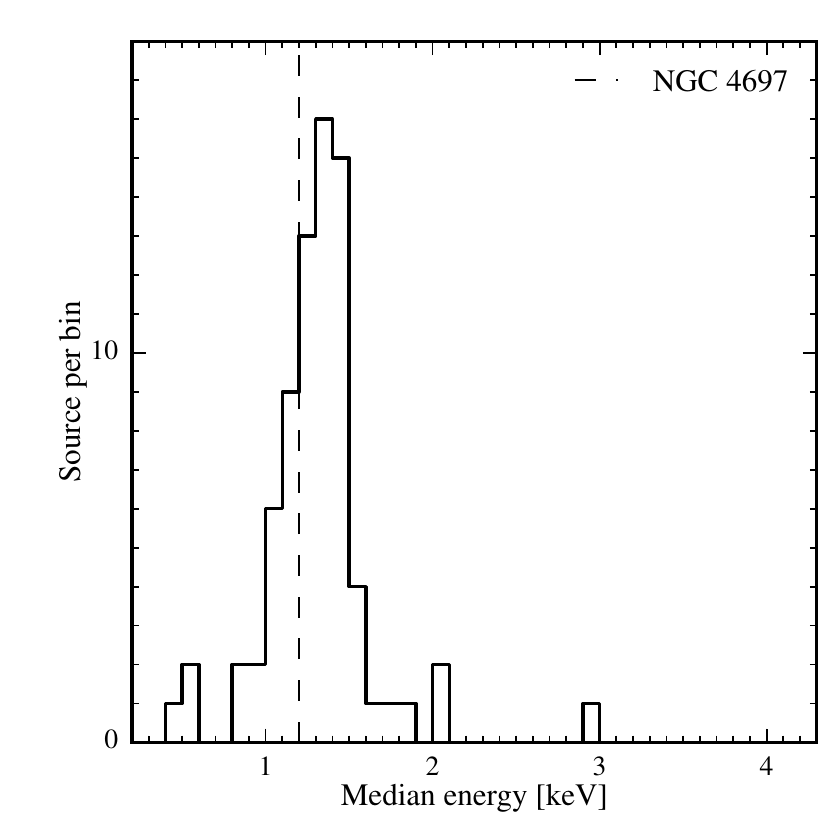}
}
\caption{Hardness ratio versus median energy diagram for  compact X-ray sources in galaxies of our sample. Only sources with  $\geq20$ net counts are plotted. Dashed lines show the median energy boundary of $\tilde{E}_0=1.2$ keV (see Section \ref{sec:emed-hr}). Distributions of sources over the median energy are plotted below corresponding median energy -- hardness ratio  diagram.}
\label{fig:HR_MED}
\end{center}
\end{figure*}

The distribution of sources on the hardness ratio -- median energy plane is shown in Fig.\ref{fig:HR_MED}. In producing these plots we applied to all data sets cuts of 20 net counts in order to exclude  sources with too large uncertainties. For each galaxy we also plot  distribution of sources over the median energy.

\subsubsection{Late-type galaxies}

Distributions of the sources over median energy in M51, M83 and M101 are clearly bimodal. Their HR-$\tilde{E}$ diagrams have a distinct and relatively compact populations of soft sources at $\tilde{E}\la 1$ keV and $HR\sim 0$ along with the more extended tail of sources with more usual harder spectra. Based on their distributions over the median energy we chose the boundary of  $\tilde{E}_0 = 1.2$ keV to select soft sources in these galaxies.  Hard sources  were excluded from further consideration. We also  excluded  from further analysis sources having statistically significant flux ($\ge 3\sigma$) detected above 2 keV. Total numbers of so selected  soft  sources  are presented  in Table \ref{tab:short_obs}.

The median energy distribution of  sources in M81 shows a single  peak at $\tilde{E}_0\approx1.5$ keV. with a possible shoulder towards lower values of $\tilde{E}$. Lack of bi-modality may be a consequence of the star-formation history of this galaxy  as discusses in section \ref{sec:Nature}.  Due to absence of obvious bi-modality  of the source distribution we applied selection criterion  based on the absence of hard flux. In particular, we excluded from further analysis sources, having statistically significant flux ($\ge 3\sigma$) detected above 2 keV. Their numbers are listed  in Table \ref{tab:short_obs} .

\subsubsection{Early-type galaxies}
Distributions of sources over median energy for elliptical and lenticular galaxies is also uni-modal with the low energy shoulder present for some galaxies.  Similar to  M81, we kept only sources without statistically significant flux above 2 keV.

\section{Observed properties of soft sources}

\subsection{Equivalent hydrogen column density}
\label{sec:nh}

\begin{table*}
\caption{Absorption column densities for late-type galaxies.
\label{tab:Nh_colums}}
\renewcommand\arraystretch{1.5}
\centering
\begin{tabular}{lcccl}
\hline
Name                    &  N$_{\rm H,f=0.5}$       &  N$_{\rm H,f=1}$         & Metallicity           & References\\
                              & (10$^{21}$ cm$^{-2}$)  & (10$^{21}$ cm$^{-2}$)     &  (Z$_\odot$)     &for metallicity\\
                     (1)     &                               (2)   & (3)                                    & (4)                 & (5)\\
\hline
NGC 5236 (M83)   &3.35  &6.35         & $\sim$ 0.5$^{*}$   & \citet{2009ApJ...695..580B}\\
NGC 5194 (M51)   & 1.84 &  3.51       &$\sim$ 0.7$^{*}$    &  \citet[][see Table 8,KK04 calibration]{2010ApJS..190..233M}\\
NGC 3031 (M81)   &1.30  & 1.53        & $\sim$ 0.7$^{*}$    & \citet{2014AA...567A..88S}\\
NGC 5457 (M101) &1.67  &2.46        & 1                            & Assumed for spectral fit\\
\hline
\end{tabular}
\flushleft
(1) -- Galaxy name. (2) and (3) -- Median value of N$_{\rm H}$ of  soft sources calculated assuming $f=0.5$ and $f=1$ (eq.\ref{eq:nh_total})  respectively.  (4) -- Metallicities of galaxies used in spectral fit and (5) -- references for these values.
Metallicities  marked with (*) were recalculated relative to solar abundance [O/H]$_{\odot}$= 4.9$\times10^{-4}$ of \citet{2000ApJ...542..914W}.
\end{table*}

\begin{figure*}
\begin{center}
\hbox
{
\includegraphics[width=60mm]{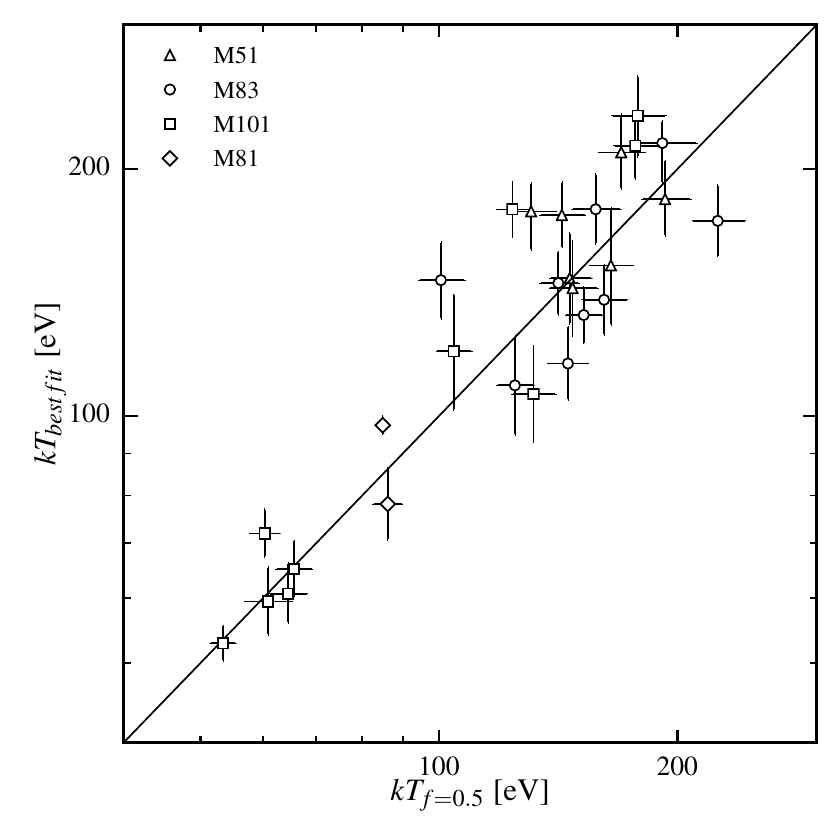}
\includegraphics[width=60mm]{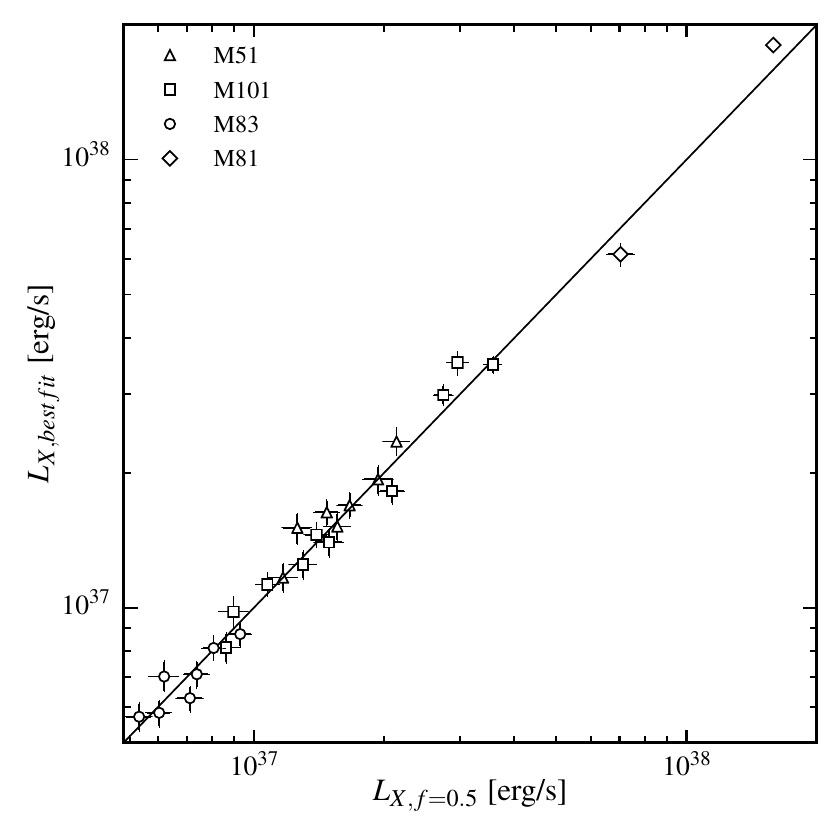}
\includegraphics[width=60mm]{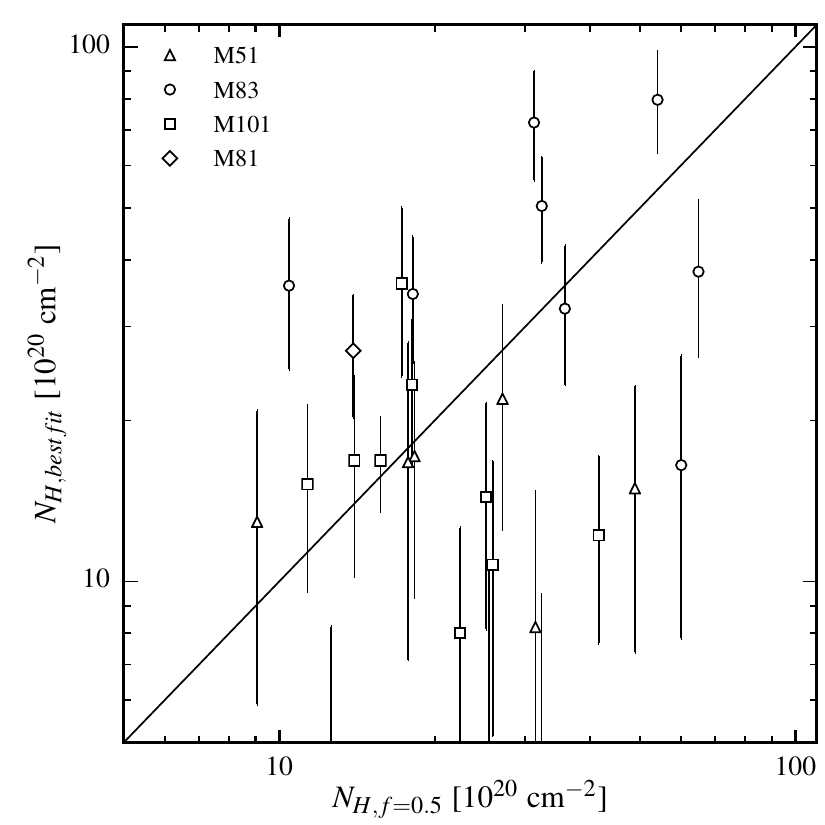}
}
\caption{
Relations between black body temperature (left) and absorbed X-ray luminosity (centre) values computed with N$_{\rm H}$ being a free parameter of X-ray spectral fit (y-axes) and for N$_{\rm H}$  computed from Eq.\ref{eq:nh_total} with $f=0.5$ (x-axes). The right hand panel shows relation between bets-fit N$_{\rm H}$  and N$_{\rm H}$  computed from Eq.\ref{eq:nh_total}. Only  sources  with net counts $\ge$200 are shown. Errors are $1\sigma$. See Section \ref{sec:nh} for more details.
 }
\label{fig:NH_fit}
\end{center}
\end{figure*}

Due to  the softness of  sources under consideration, the value of equivalent hydrogen column density $N_H$ is important for accurate spectral analysis. This problem is of relevance for late-type galaxies.

Because of degeneracy with other parameters $N_H$  is difficult to determine  from  fitting X-ray spectra, especially for fainter sources with smaller number of counts, which constitute  the majority of our sample. Therefore we chose to fix $N_H$ at values determined in radio band. To this end we combined CO 2-1 and 21 cm maps from  the HI Nearby Galaxy Survey (THINGS, \citealt{2008AJ....136.2563W}) and the HERA CO-Line Extragalactic Survey (HERACLES,  \citealt{2009AJ....137.4670L}). For each source we computed absorption column density as
\begin{equation}
N_{\rm H}=N_{\rm H,Gal}+f\times \left(N_{\rm HI}+2\times N_{\rm H_2}\right)
\label{eq:nh_total}
\end{equation}
where $N_{\rm H,Gal}$ is the Galactic hydrogen column density in the direction of galaxy \citep{2016A&A...594A.116H},  $N_{\rm HI}$ and $N_{\rm H_2}$ are column densities of neutral and molecular gas from THINGS and HERACLES surveys respectively, computed as follows:

\begin{equation}
N_{\rm HI} [\rm cm^{-2}]=\frac{1.11\times 10^{24}\times \rm I_{HI}}{FWHM_{\rm maj}[\rm \arcsec] \times FWHM_{\rm min}[\rm \arcsec]}
\label{eq:nh1}
\end{equation}
where   I$_{\rm HI}$ is velocity-integrated intensities of THINGS maps in units of Jy beam$^{-1}$ km s$^{-1}$.  FWHM$_{\rm maj}$ and FWHM$_{\rm min}$ are the major and minor axes of the beam in arsec \citep[][see Eq. 5]{2008AJ....136.2563W}. For $N_{\rm H_2}$ we used:
\begin{equation}
N_{\rm H_2}[\rm cm^{-2}] =\frac{R_{\rm 21}}{0.8} \times X_{\rm CO}\times W_{\rm CO},
\label{eq:nh2}
\end{equation}
where W$_{\rm CO}$ is HERACLES velocity-integrated CO J = 2 $\rightarrow$ 1 map intensity in units of K km s$^{-1}$,  X$_{\rm CO}$ is the CO J = 1 $\rightarrow$ 0 to H$_{\rm 2}$ conversion factor and R$_{\rm 21}$ is CO J = 2 $\rightarrow$ 1 to CO J = 1 $\rightarrow$ 0 ratio. We assumed Galactic value of X$_{\rm CO}=$ 2$\times$10$^{20}$ cm$^{-2}$ (K km s$^{-1}$)$^{-1}$  \citep[][]{1996A&A...308L..21S,2001ApJ...547..792D} for majority of  galaxies and R$_{\rm 21}$=0.8 \citep[see][Eq.3 and references therein]{2009AJ....137.4670L}.

The  X$_{\rm CO}$ factor is known to vary between galaxies, in particular for M51 it is $\sim4$ times smaller than the Galactic value \citep{1993A&A...274..123G,1995A&A...298L..29G}. For M81, no CO maps of sufficient quality are available and we accounted only for absorption by neutral hydrogen.
For M83 we used CO 1--0 intensity map from NED\footnote{NASA/IPAC Extragalactic Database: https://ned.ipac.caltech.edu}  where galaxy was observed by \citet{2002AJ....123.1892C} with NRAO 12m telescope.

Table \ref{tab:Nh_colums} shows median values of absorption column densities $N_{\rm H}$ of soft sources in late-type of galaxies for two cases: $f=0.5$ and $f=1$. We adopted $f=0.5$ i.e. that sources are subjected to a half of the absorption intrinsic to galaxy. Naively, this corresponds to the assumption that  all sources are located close to the mid-plane of the host galaxy.

We analysed spectra of bright sources with  $\ge$200 net counts and compared results of X-ray spectral fits with fixed $N_H$ with spectral fits where $N_H$ was a free parameter of the fit. The results are shown in Fig.\ref{fig:NH_fit}. As one can see, although there is a considerable scatter in the $N_H$ values, values of temperature and, especially, X-ray luminosity are in a reasonable overall agreement. Unfortunately, such analysis is possible only for a small fraction of sources. We therefore chose to use fixed absorption column density, computed, for each source, from Eq.\ref{eq:nh_total}.

Elliptical and lenticular galaxies in our sample  have much lower intrinsic neutral hydrogen column density,  below $\rm \sim {\rm few} \times10^{19}\ cm^{-2}$ \citep{2012MNRAS.422.1835S}. We therefore can safely assume  $N_H=N_{\rm H,Gal}$ for early-type  galaxies in our sample, with a caveat regarding  Centaurus A.  In the dust lanes of this galaxy, absorption column density  is at the nearly constant level of  $\approx 5\times 10^{21}$ cm$^{-2}$ and drops rather sharply outside the dust lanes \citep{2010A&A...515A..67S}. Therefore, for the sources located in the dust lanes (about $\sim 1/3$ of all soft sources in Centaurus A) we accounted for the intrinsic absorption in the same way as it was done for spiral galaxies, but assuming a constant value of intrinsic absorption of $5\times 10^{21}$ cm$^{-2}$, while for the sources outside dust lanes we used Galactic value. We found that particular details of this calculation are unimportant. It should be also noted that Centaurus A galaxy also stands out in the early-type galaxy  sample  due to its ongoing star-formation on the level of $\sim 0.8 ~M_\odot$/yr  \citep{2019ApJ...887...88E}. It will be further discussed  in Section \ref{sec:SSS_frequency}.

\subsection{X-ray spectral analysis and $kT_{bb}-L_X$ diagram}
\label{sec:temp}

\begin{figure*}
\begin{center}
{
\includegraphics[width=60mm]{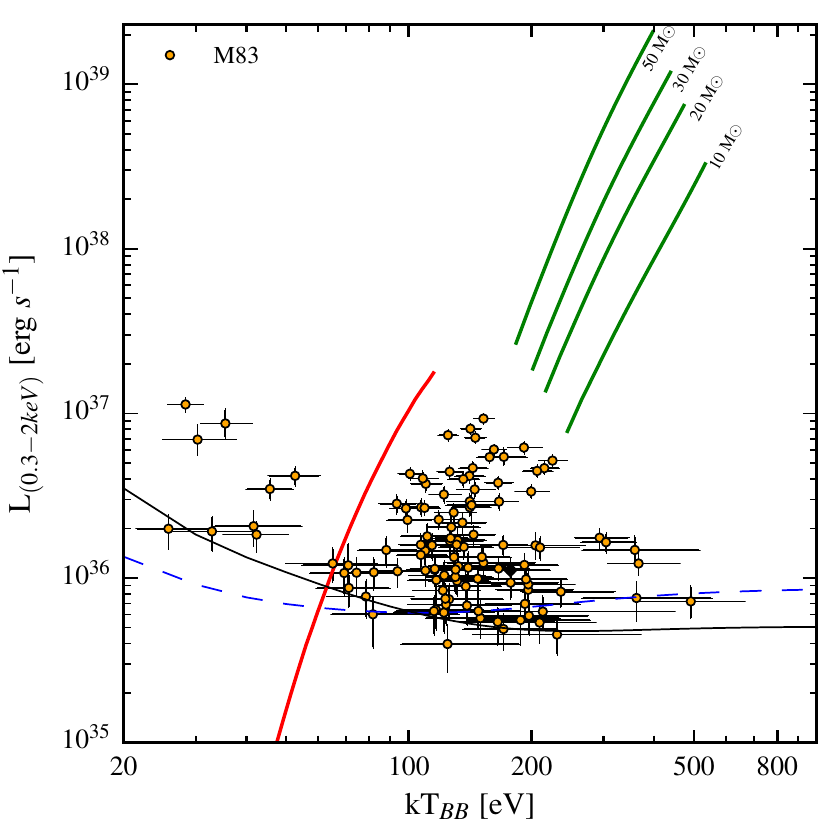}
\includegraphics[width=60mm]{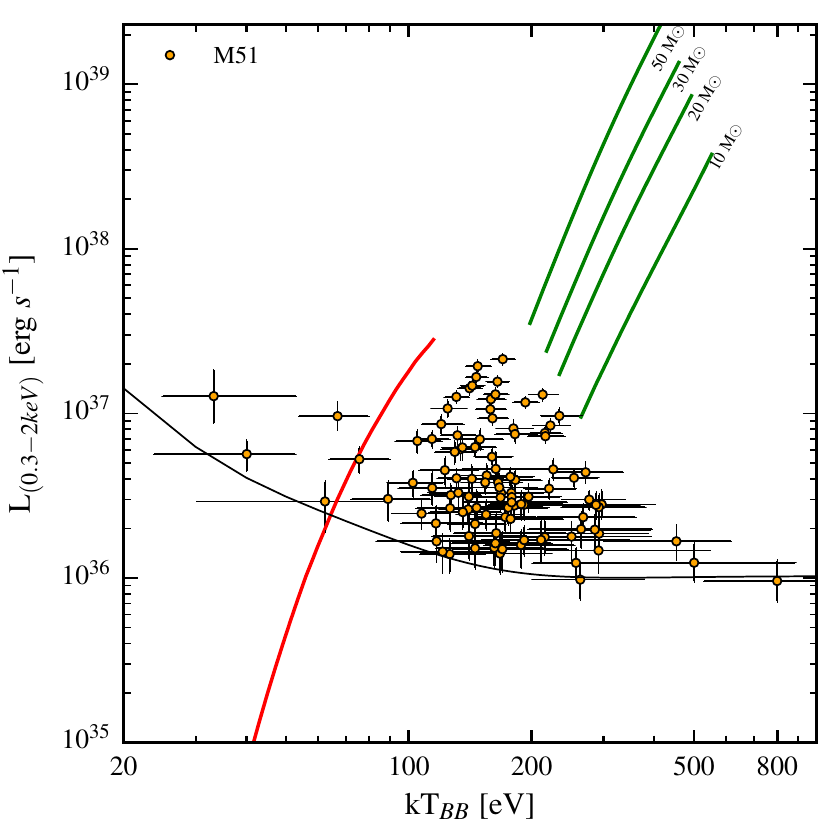}
\includegraphics[width=60mm]{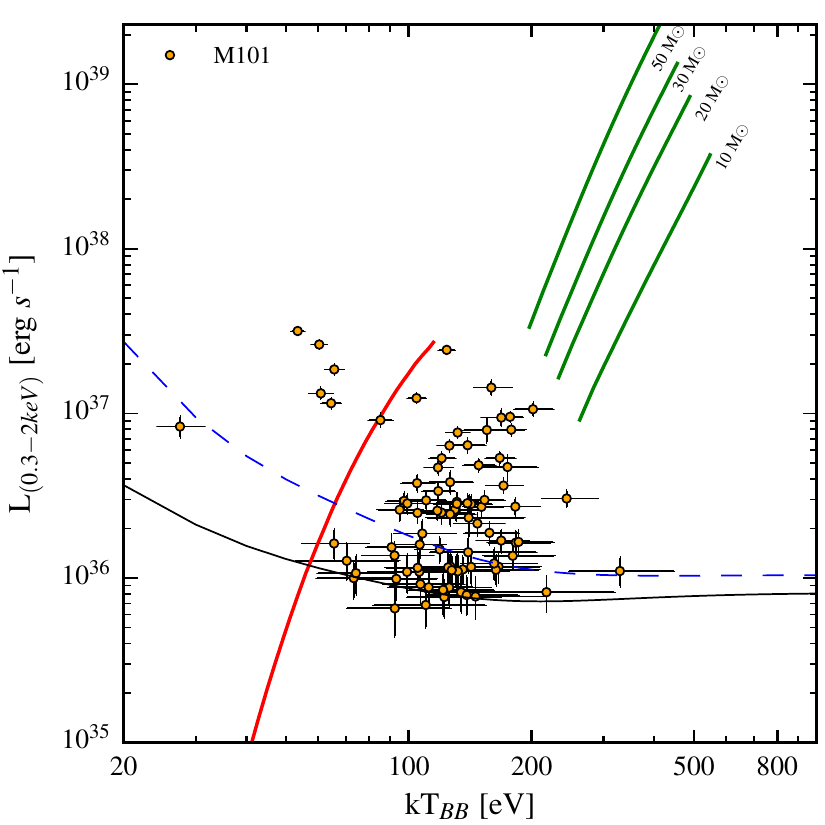}
}
\hbox
{
\includegraphics[width=60mm]{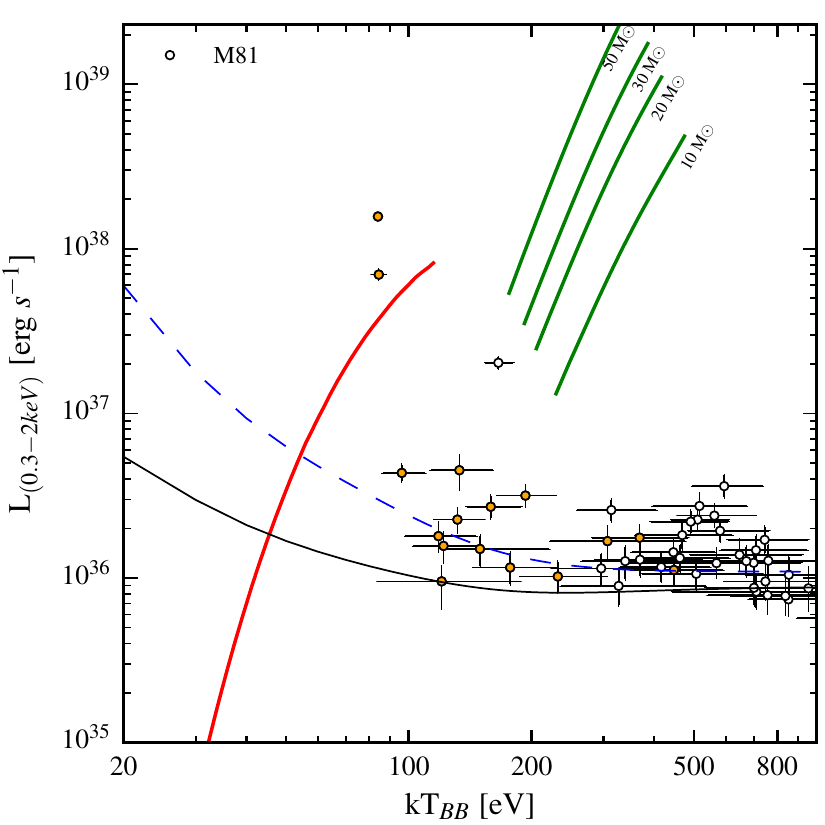}
\includegraphics[width=60mm]{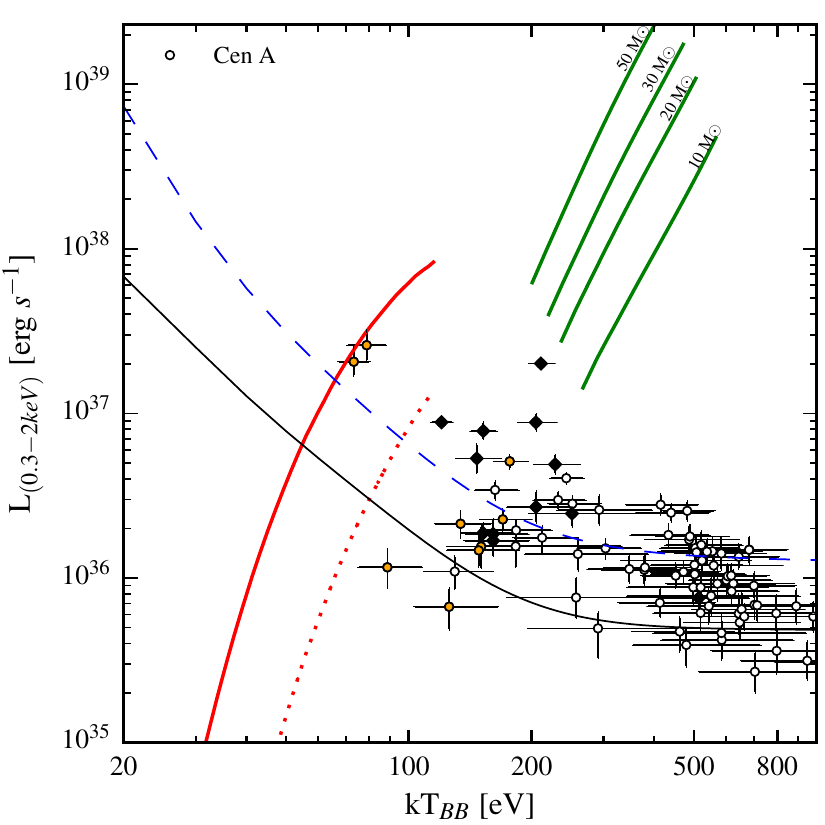}
\includegraphics[width=60mm]{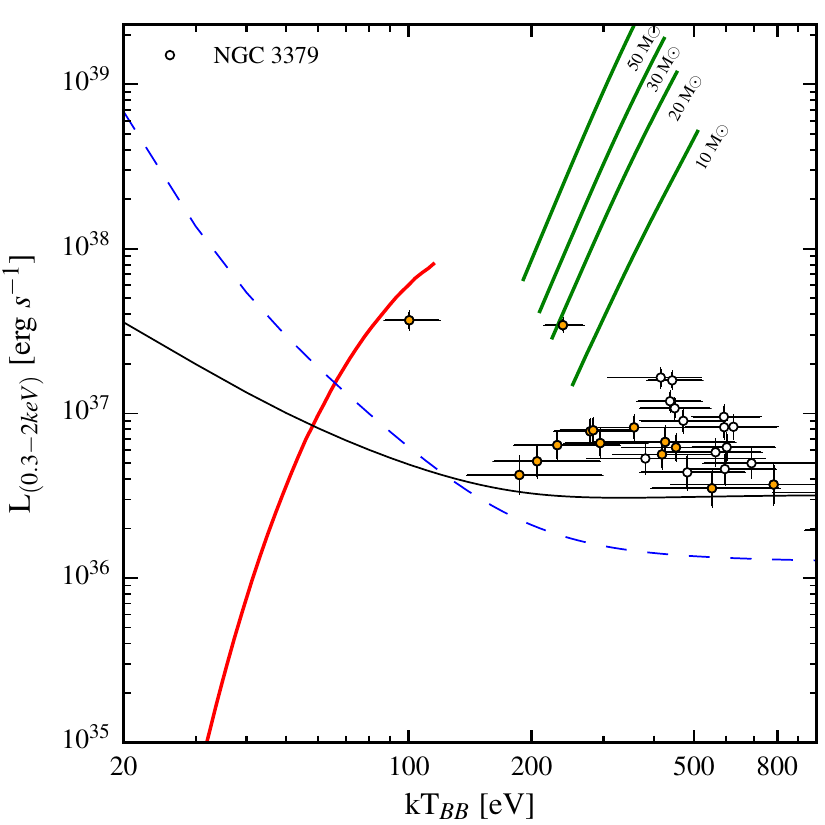}
}
\hbox
{
\includegraphics[width=60mm]{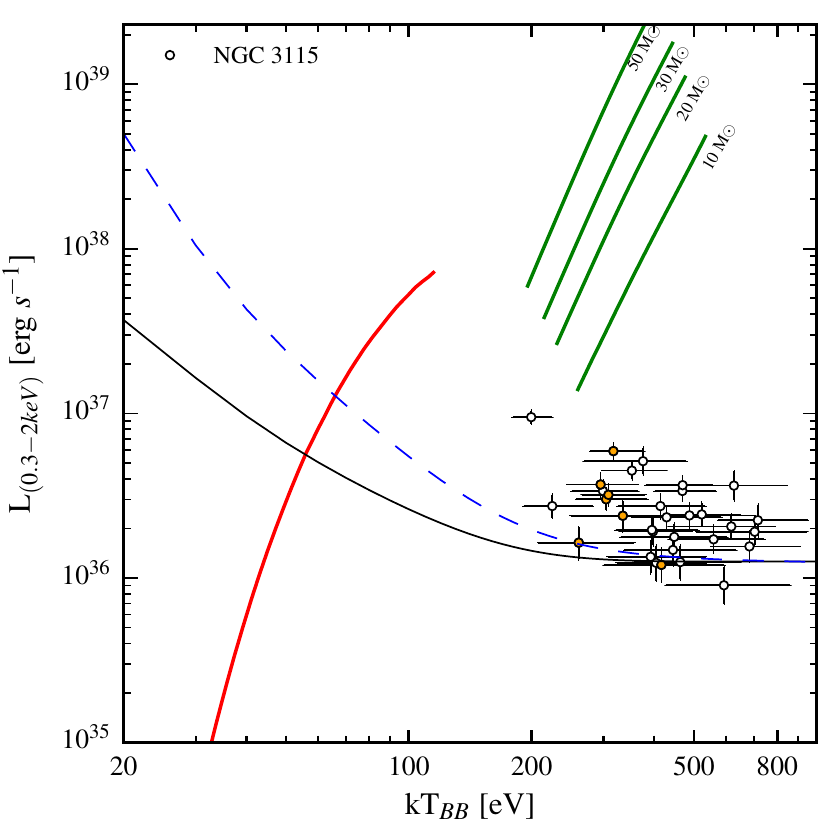}
\includegraphics[width=60mm]{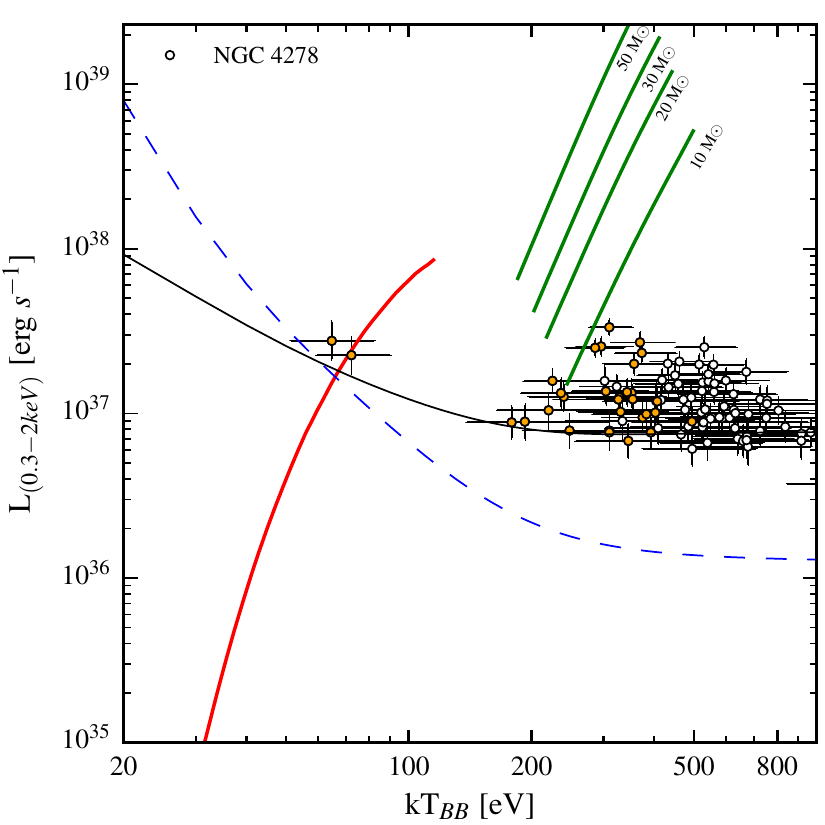}
\includegraphics[width=60mm]{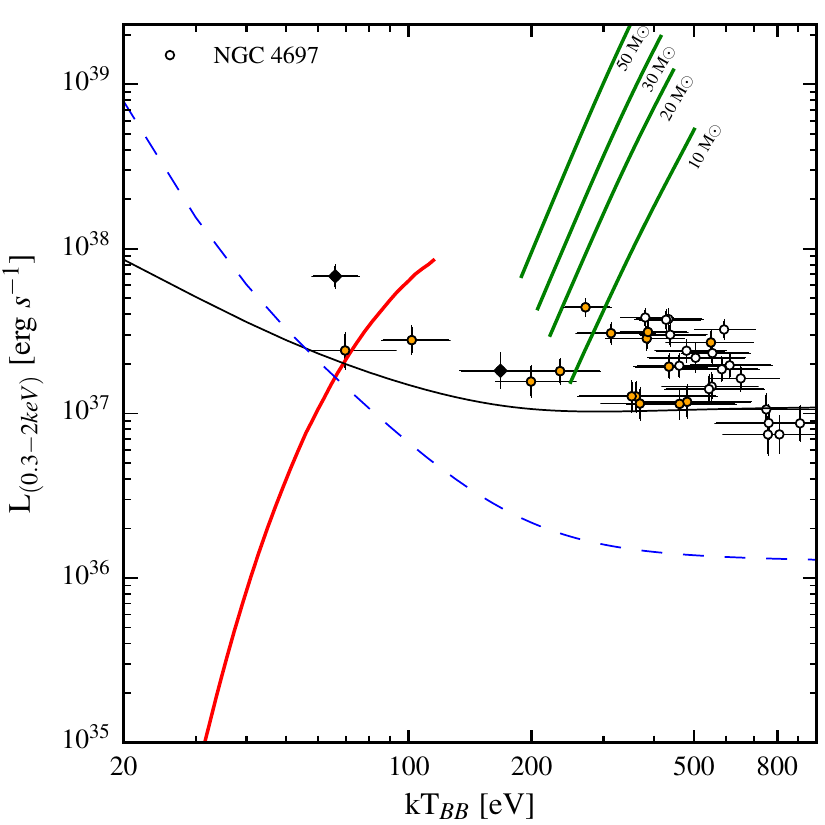}
}
\end{center}
\caption{The $kT_{\rm bb}-L_{\rm X}$ diagram of soft X-ray sources in galaxies from our sample. A cut of 20 net counts was applied. The luminosities are in the $0.3-2.0$ keV band and are not corrected for absorption. All plotted sources have no statistically significant flux above 2 keV  ({\it open circles}). Sources with median energy $\tilde{E} \le 1.2$ keV are shown with {\it solid orange circles}.  Sources, having Gaia matches  with measured parallax or proper motion are  marked by  {\it black diamonds}. {\it Solid red curves} show $kT_{bb}-L_X$ relations  at the lower boundary of stable nuclear burning. Along these curves, the WD mass increases from 0.51$M_{\sun}$ at the low-left end to 1.34$M_{\sun}$ at the upper right end. The dotted red line for Centaurus A shows location of this boundary in the dust lanes of this galaxy.  {\it Green lines} correspond to the emission spectrum of general relativistic accretion disk model around a Schwarzschild black hole \citep{1991ApJ...367..213E} for BH masses of 10, 20, 30 and 50$M_{\sun}$. The accretion rate changes along the curves from $0.02  \dot{M}_{\rm Edd}$ to $ \dot{M}_{\rm Edd}$. Horizontally oriented {\it solid black} curves show the luminosity, corresponding to 20 net counts. {\it Dashed blue line} on each plot shows the  20 counts luminosity limit for M51 corrected for the difference in absorption column density between M51 and the galaxy under consideration.  All model curves show absorbed luminosities. Errors are at $1\sigma$ confidence. See Section \ref{sec:temp} for details of calculations.
}
\label{fig:lum_kt}
\end{figure*}

Spectra of X-ray sources, classified as soft, were approximated with an absorbed black body model, with the absorption column density being fixed at the values computed as described in Section \ref{sec:nh}. For spectra with small number of counts absorption correction of X-ray luminosity often bears large uncertainties and may lead to physically meaningless values. Therefore we chose to use absorbed X-ray luminosities of  sources in further analysis and apply absorption to  models, when comparing them to the data.  We used Tuebingen-Boulder ISM absorption model \citep{2000ApJ...542..914W}  ({\it tbvarabs} model in XSPEC).
Abundances of late-type  galaxies were fixed at values listed on Table \ref{tab:Nh_colums}. For early-type  galaxies \citet{{2000ApJ...542..914W}} solar elemental abundance was assumed.
The spectral fit was performed with  {\it XSPEC}  \citep{1996ASPC..101...17A} in the 0.3--8 keV energy band using C -- statistics \citep{1979ApJ...228..939C}. Background spectrum was included in the spectral fitting. Spectra were binned to have at least one count per bin, following recommendations of the XSPEC manual\footnote{https://heasarc.gsfc.nasa.gov/xanadu/xspec/\\ manual/XSappendixStatistics.html}. Confidence intervals for the parameters  were obtained using {\it error} command in XSPEC.

Distribution of sources on the X-ray luminosity -- temperature (hereafter $kT_{\rm bb}-L_{\rm X}$) plane  is shown in Fig. \ref{fig:lum_kt}. In plotting $kT_{\rm bb}-L_{\rm X}$  diagram we applied a cut of 20 net counts, in order to exclude faint  sources with too large error bars. The X-ray luminosities were computed for the  0.3--2 keV energy band.
The thin solid line near the bottom of each panel shows X-ray luminosity  corresponding to 20 net counts, as a function of the black body temperature. In plotting these curves we converted counts to rate using the maximal exposure time of the observation. The absorbed X-ray luminosity was computed using count rate -- flux conversion for the black body spectrum with the given temperature and absorption column density. The  luminosity corresponding to 20 counts exceeds by a factor of $\sim$3 luminosities of  faintest sources, therefore  samples of sources above these curves should not be subject to significant incompleteness effects.\footnote{See, for example, incompleteness function of Chandra data for Centaurus A in \citet{2006A&A...447...71V}. From their Fig A1, incompleteness effects become relatively small  at luminosities  $\sim 2-3$ larger than detection limit of the data.}

On each panel, the blue dashed line shows the luminosity limit for M51 data corrected for the difference between absorption column densities in M51 and the galaxy under consideration. The correction factor  is close to unity  at temperatures exceeding a few hundred  eV. However, it may be considerable at low temperatures, for soft spectra.
These lines permit to compare sensitivities of different datasets independently of the absorption. In particular, this comparison demonstrates that despite the rather significant scatter in Chandra exposure and distances to galaxies,  selected galaxies have comparable  detection sensitivity to sources with soft spectra.

In Fig.\ref{fig:lum_kt},  solid red curves show the lower bound of the stability strip for the nuclear burning of hydrogen  on the WD surface  \citep{2013ApJ...777..136W}.  In plotting these curves,  we used the Stefan-Boltzmann law to compute the effective temperature and the mass-radius relation for zero-temperature white dwarfs from \citet{2000A&A...353..970P}.  The WD mass changes along the curve  from 0.51$M_{\sun}$ to 1.34$M_{\sun}$.
We then used the black body model in XSPEC, apply ISM absorption  and compute 0.3--2 keV band luminosity, corresponding to the given WD mass and effective temperature.  For late-type galaxies, absorption column densities were fixed at corresponding median values from Table \ref{tab:Nh_colums} while for  early-type galaxies we used the Galactic value $N_{\rm H,Gal}$, as explained above. For Centaurus A we also show location of this boundary for sources located in the dust lanes. Green lines in Fig.\ref{fig:lum_kt} show tracks of the spectral model of relativistic accretion disk  around a Schwarzschild black hole \citep{1991ApJ...367..213E}.  Along these lines, the mass accretion rate changes  from $0.02 \dot{M}_{\rm Edd}$ to $\dot{M}_{\rm Edd}$, different lines correspond to the different black hole mass as marked by the number near the curves.  To compute these tracks, we  approximated the accretion disk emission spectrum by an absorbed black body model.  The absorption was applied similar to other  models plotted in the figure.

All sources plotted in Fig.\ref{fig:lum_kt} have no significant ($\ge 3\sigma$) flux above 2 keV.  For three late-type galaxies  with bi-modal median energy distributions (M83, M51 and M101) we plot  only sources with median energy $\tilde{E}\la 1.2$ keV. These sources  show rather compact  distributions in the $kT_{\rm bb}-L_{\rm X}$ plane with temperatures 20$\sim$300 eV and X-ray luminosities in 10$^{36}$$\sim$10$^{38}$ ergs/s range. For other galaxies, we only used the criterion of lack of statistically significant ($\ge 3\sigma$) emission above 2 keV. For these galaxies, distributions of sources extend significantly towards higher temperatures (cross-hair symbols in Fig.\ref{fig:lum_kt}), which is not surprising as such a selection includes many sources with regular spectra. These sources are dominated by X-ray binaries and background AGN and their discussion is beyond the scope of this paper.   Sources with low median energy $\tilde{E}\la 1.2$ keV (solid orange symbols) tend to have lower temperatures than the bulk of sources in these galaxies and will be discussed in the following section, along with soft sources in M83, M51 and M101.

\subsection{Foreground stars}
\label{sec:stars}

To identify foreground stars we cross-correlated the lists of soft  X-ray sources with the GAIA \citep{2016A&A...595A...1G} catalog, Data Release 2 \citep{2018A&A...616A...1G}. With the search radii of $2 \arcsec$ we typically found $\sim$2 matches per galaxy except for  Centaurus A galaxy where we found 12 matches. Large number of matches for Centaurus A is caused by its relative proximity to the Galactic plane. Some of the GAIA matches  have  statistically significant  measurements of parallax or proper motion, with S/N>3. These sources are classified as stars in the Galaxy and excluded from further analysis. We mark them with open diamonds in Fig. \ref{fig:lum_kt}.

\section{Nature of soft and super-soft X-ray sources}

Based on first Chandra studies of nearby galaxies it was proposed to divide soft sources in external galaxies into super-soft and quasi-soft \citep[][]{2003ApJ...592..884D,2004ApJ...609..710D}. The former are characterised by color temperatures of $kT\la 100$ eV and are believed to be mostly associated with accreting white dwarfs with nuclear burning of hydrogen on their surface, while the latter have temperatures in the range  $kT\sim 100- 300$ eV and, possibly, are unrelated to white dwarfs.

\subsection{Super-soft X-ray sources and accreting white dwarfs}
\label{sec:SSS_sample}

In the commonly used albeit (over)simplified picture, it is assumed that  the spectral energy distribution of nuclear burning WDs can be roughly described by a black body spectrum \citep[cf.][]{woods13} with the color temperature close to the effective temperature. Observations of  well-known super-soft X-rays sources in the Milky Way, Magellanic Clouds \citep[e.g.][]{1996LNP...472..299G,1996IAUS..165..425K} seem to broadly support this assumption. The effective temperature of an accreting white dwarf powered by nuclear burning of hydrogen on its surface is given by:
\begin{eqnarray}
T_{\rm eff}=\left(\frac{\epsilon_H X_H \dot{M}}{4\pi R_{\rm phot}^2 \sigma_{SB}}\right)^{1/4}=\nonumber\\
67
\left(\frac{\dot{M}}{5\times 10^{-7} ~M_\odot/yr} \right)^{1/4}
\left(\frac{R_{\rm phot}}{10^{-2} ~R_\odot} \right)^{-1/2}
~eV
\label{eq:kteff}
\end{eqnarray}
where $\epsilon_H=6\times 10^{18}$  erg/g is the energy release in hydrogen fusion per unit mass of hydrogen, $X_H\approx0.72$ is hydrogen mass fraction of the accreting material, $\dot{M}$ is the mass accretion rate and $R_{\rm phot}$ is the photospheric radius of the hydrogen burning layer. Numerical calculations show that at the bottom of the stability strip  $R_{\rm phot}\sim R_{\rm WD}$ and photospheric radius increases with increase of $\dot{M}$ \citep{2013ApJ...777..136W}.

The red solid line in Fig.\ref{fig:lum_kt} shows the temperature--luminosity relation  computed  at the lower bound of the stability strip and assuming $R_{\rm phot}=R_{\rm WD}$. For the definition of the (white dwarf mass dependent) lower boundary of stable nuclear burning we used results of \citet{2013ApJ...777..136W}  and for the WD radius we used zero temperature radius (depending on the WD mass) from \citet{2000A&A...353..970P}. The region on the $kT_{bb}-L_X$ diagramm above and to the left of  this line is permitted for accreting white dwarfs stably burning hydrogen on their surface, while in the part of the diagram  to the bottom-right stable nuclear burning of hydrogen is not possible and it should be populated with other types of soft X-ray sources.
Indeed, in Fig.\ref{fig:lum_kt}  one can see that in late-type galaxies (M83, M51, M101 and M81), there is a distinct group of rare soft and luminous sources located above these line. These sources have temperatures of $\sim$several tens of eV  and seem to  clearly stand out  in the populations of soft sources in these galaxies. In interpreting Fig.\ref{fig:lum_kt} one should bear in mind that it shows absorbed luminosities in $0.3-2.0$ keV band and that absorption and bolometric corrections  increase $\sim$exponentially with  decrease of the color temperature, therefore the softest sources are truly luminous.

Thus, we classify sources to the left of the red line in Fig.\ref{fig:lum_kt} as classical super-soft X-ray sources. We also include those sources which  are located outside the stable nuclear burning boundary but their error bars cross this boundary.
In addition, we added to this list  two sources in M101, located outside but close to the stable nuclear burning boundary (the two sources in Fig.\ref{fig:lum_kt} upper right panel with $kT\ga 100$ eV and $L_X\ga 10^{37}$ erg/s.). The catalog of super-soft X-ray sources in presented in Table \ref{table:catalog}, and their numbers in galaxies from our sample are listed in  Table \ref{tab:short_obs}.

\subsection{Specific frequency of super-soft X-ray sources across morphological types}
\label{sec:SSS_frequency}

From Fig.\ref{fig:lum_kt}, it is evident that super-soft X-ray sources are considerably more abundant in late-type galaxies than in early-type galaxies. However, for a statistically sound conclusion one needs to take into consideration completeness limits of various data sets. To this end we used the sensitivity limit for M51 galaxy,  corrected for the difference  in the absorption column density as described in Section  \ref{sec:temp}, to select super-soft X-ray sources.  In each galaxy, we selected sources to the left of the thick red line and above the dashed blue line in Fig. \ref{fig:lum_kt}.   This selection is within  the 20 counts  flux limit (solid black line in Fig. \ref{fig:lum_kt}.)  in all galaxies,   except for  M83, where at the low temperature end, the 20 counts limit for this galaxy  is by a factor of about $\sim 2$ higher. This may result to slight underestimation of the specific frequency of super-soft sources in late-type galaxies. In counting super-soft sources, we included all the sources which error bars crossed the stable nuclear burning boundary. We also included two sources in M101, located outside but close to the stable nuclear burning boundary (the two sources in Fig.\ref{fig:lum_kt} upper right panel with $kT\ga 100$ eV and $L_X\ga 10^{37}$ erg/s).  In Centaurus A galaxy, we used different criteria for sources locate in and outside the dust lanes (c.f. dotted line in Fig. \ref{fig:lum_kt}), although this does not change the final result.

In order to estimate specific frequencies of super-soft X-ray sources, we group the data separately for early- and late-type galaxies. There are 29 super-soft sources in late-type galaxies (M83, M51, M81, M101), which, for the total stellar mass of  13.9$\times10^{10}$ M$_{\odot}$  yield the specific frequency of
\begin{equation}
f_{\rm SSS}^{\rm S}\approx(2.08\pm0.46)\times10^{-10}~ M_{\odot}^{-1}
\end{equation}
In computing statistical errors we used  Gehrels approximation \citep{1986ApJ...303..336G}.
Counting lenticular  and elliptical galaxies together (Centaurus A, NGC3115, NGC3379, NGC4278, NGC4697), we find 7 super-soft X-ray sources sources for the total stellar mass of $28.3\times10^{10}$ M$_{\odot}$ yielding specific frequency of
\begin{equation}
f_{\rm SSS}^{\rm E+S0}\approx (2.47\pm1.34)\times10^{-11} M_{\odot}^{-1}
\end{equation}
Centaurus A galaxy, although classified as S0, shows some star-formation activity  within its dust lanes, at the  the level of $\sim 0.8 ~M_\odot$/yr  \citep{2019ApJ...887...88E}. However, its SSS specific frequency is closer to early-type galaxies  than to late-type galaxies. In order to check how its inclusion in our sample affects  our results, we compute SSS specific frequency for elliptical galaxies only:
\begin{equation}
f_{\rm SSS}^{\rm E}\approx (2.02\pm1.38)\times10^{-11} M_{\odot}^{-1}
\end{equation}

This number is $\sim 20\%$ smaller but compatible with the value obtained including Centaurus A.

The lower limit for the SSS specific frequency in early-type galaxies at the $3\sigma$ confidence level is
$f_{\rm SSS}^{\rm E}> 8.2\times10^{-12} M_{\odot}^{-1}$.

The ratio of specific frequencies of SSS in late-type to early-type galaxies is
\begin{equation}
f_{\rm SSS}^{\rm E+S0}/ f_{\rm SSS}^{\rm S}=0.12\pm0.05
\end{equation}
i.e. late-type galaxies contain about $\approx 8$ times more super-soft X-ray sources than early-type galaxies. Using Bayesian approach, we estimate that  their  specific frequencies  are different at the confidence level corresponding to about $\approx 5.8\sigma$.

We searched for variations of the SSS specific frequency between galaxies of the same morphological type. To this end we compared observed numbers of super-soft X-ray sources in each galaxy with their expected values  computed from the corresponding SSS specific frequency and stellar mass of the galaxy.  For early-type galaxies  we found  no statistically significant deviations of observed numbers of super-soft X-ray sources from their expected values.  This is not surprising, given small numbers of SSS in early-type galaxies. For the  late-type galaxies, however, we observe statistically significant deviations in M83 and M81 where the observed numbers of super-soft sources differ from the expected  values  at the statistically significant levels exceeding $\sim 3\sigma$. This can be understood considering that populations of accreting nuclear burning white dwarfs are determined by the long-term star-formation history of the galaxy, rather than by its mass, as discussed in the following section.

For late-type galaxies, we also compute the SSS specific frequency per unit star formation rate.  Using SFR values, collected from the literature (Table \ref{tab:galaxy_sample} and references there) we obtain:
\begin{equation}
f_{\rm SSS}^{\rm S}\approx(3.13\pm0.70)~ \left(M_{\odot}/yr\right)^{-1}
\end{equation}
Overall, the numbers of super-soft X-ray sources in late-type galaxies  appear to  somewhat better correlate with their star-formation rate than with the stellar mass.

\subsection{Comparison with previous work and discussion}

The fact that super-soft X-ray sources tend to be predominantly associated with young stellar populations has been noted previously \citep[e.g.][]{2004ApJ...609..710D}.  {Studying populations of soft sources in nearby galaxies, \citet{2010ApJ...712..728D}  mentioned    the difference in the numbers of soft sources  in early- and late-type  galaxies, however they did not quantify  these trends. \citet{2011MNRAS.412..401B}  using sample of 12 nearby late-type galaxies concluded  that specific frequencies (per unit K-band luminosity) of luminous, $L_X>10^{36}$ erg/s,  super-soft sources in disks of spiral galaxies by factor of $\sim$2 exceed that of bulges. Taking into account that the typical $M_*/L_K$ ratio for bulges are approximately twice larger than for disks, this translates to about 4 times difference in specific frequency per unit mass, i.e. shows qualitatively same trend as our comparison of late-type and early-type galaxies, although the particular numbers are different. The difference is due to the fact, that  \citet{2011MNRAS.412..401B} used a  selection procedure to identify super-soft X-ray sources, based on the hardness ratio with the threshold, corresponding to $kT$=200 eV.

Population synthesis calculations of nuclear  burning accreting WDs predict difference of specific frequencies of SSSs  in spiral and elliptical galaxies  \citep{2015MNRAS.453.3024C}. Using their model {\tt a025qc17} we estimate specific frequencies of super-soft X-ray sources with X-ray luminosities exceeding $10^{36}$ erg/s. For spiral galaxies they obtained  specific frequencies  ranging from  $4\times10^{-10}$ to $2\times10^{-9}$  sources per $M_\odot$, corresponding to  absorption column densities of $3\times10^{21}$ cm$^{-2}$ and $3\times10^{20}$ cm$^{-2}$. For elliptical galaxies they obtained   $\approx 6\times 10^{-11}$  sources per $M_\odot$ (for column density $3\times10^{20}$ cm$^{-2}$), i.e. $\ga 7$ times less frequent that in late-type galaxies. The contrast between late- and early- type galaxies is well consistent with our observations, especially taking into account   dependence of these numbers on the age of the host galaxy \citep{2015MNRAS.453.3024C}. However, the actual numbers of specific frequencies are by at least  a factor off $\sim  2-3$ larger in simulations than in our data. Among other reasons, this can be understood as a result of an approximate  account for  absorption, as the numbers of observable super-soft X-ray sources are highly  sensitive to the column density.

Another  interesting result of this work is that within the morphological types, the number of super-soft X-ray sources does not seem to obey  simple scaling relations with parameters of the host galaxy. In particular, there are large and statistically significant excursions in the specific frequency of SSS in spiral galaxies, normalised  to their stellar mass.  This can  be expected in the view of the results of WD population modelling which predict that  the population of super-soft sources peaks at about 1 Gyr after the star formation event \citep{2015MNRAS.453.3024C}.
For this reason,   numbers of super-soft sources in  galaxies should not be expected to  scale with stellar-mass or current star-formation rate.
In the more general context, low specific frequency of super-soft X-ray sources, especially  in early-type galaxies, support earlier conclusion that they are not the major   class of type Ia supernova progenitors \citep{2010ApJ...712..728D, 2010Natur.463..924G,johansson14, woods14}.

\subsection{Other soft X-ray sources}
\label{sec:Nature}

\begin{figure}
\centering
\includegraphics[width=0.45\textwidth,clip=true]{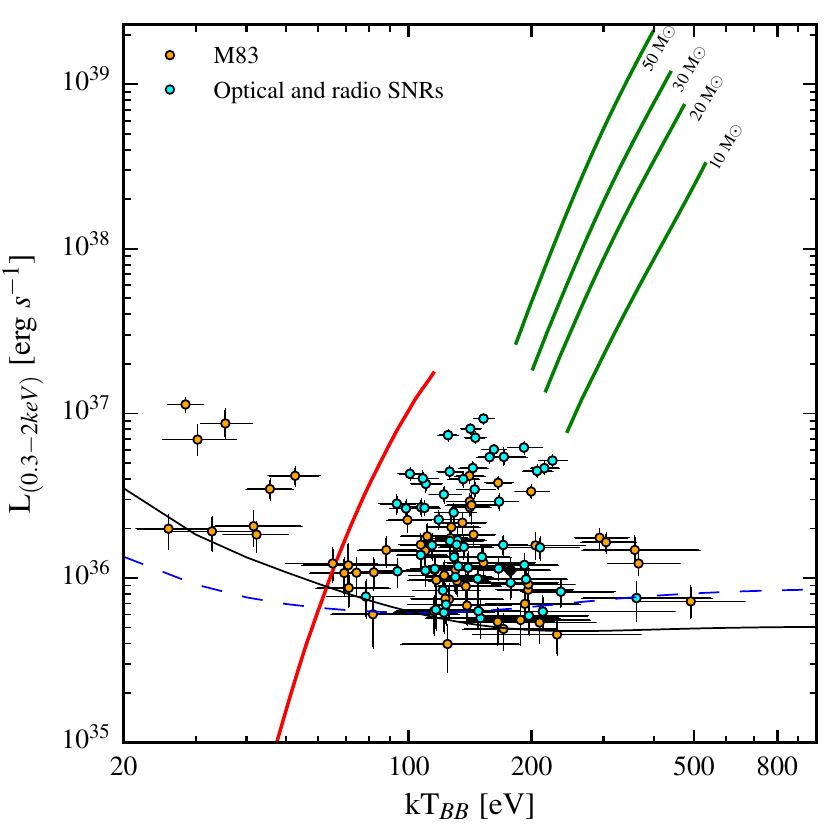}
\caption{
The $kT_{\rm bb}-L_{\rm X}$ diagram of soft X-ray sources in M83. Supernova remnants confirmed by optical and radio data  are shown by cyan circles. See  caption to Fig. \ref{fig:lum_kt}  for detailed description of  $kT_{\rm bb}-L_{\rm X}$ diagram.}
\label{fig:m83}
\end{figure}

\begin{figure}
\centering
\includegraphics[width=0.45\textwidth,clip=true]{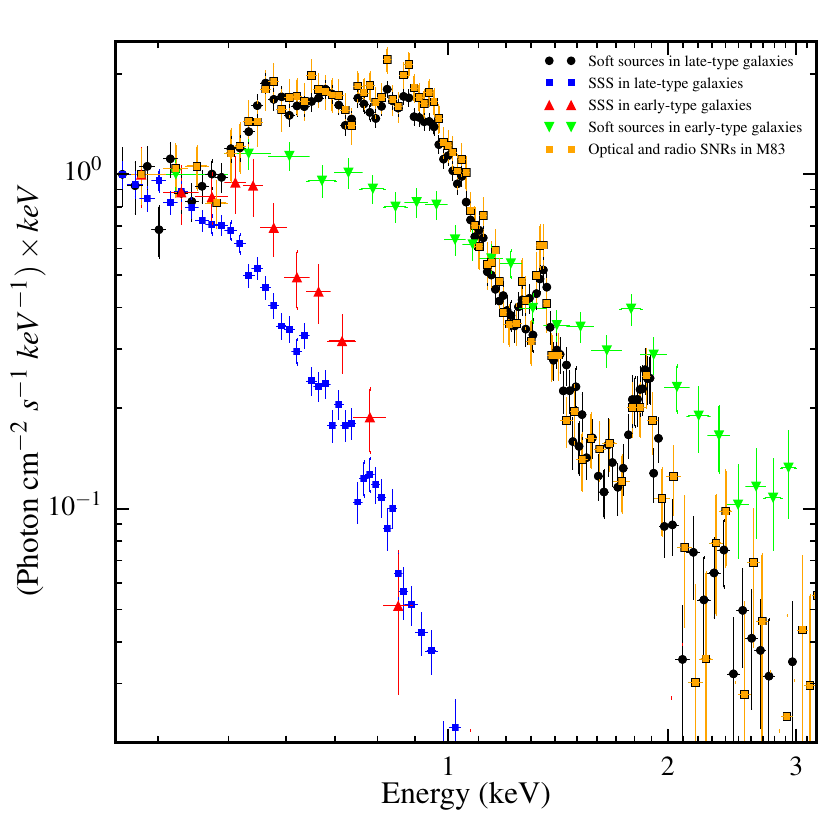}
\caption{
Combined spectra of soft and super-soft X-ray sources in late- and early-type galaxies. {\it Black and green symbols} show integrated spectrum of all soft sources located outside the stable nuclear burning region for WDs (i.e. located to the right of the thick red line in Fig.\ref{fig:lum_kt}) in late-  and early-type galaxies. {\it Yellow symbols} show  integrated spectrum of sources, matching confirmed SNRs in M83.  {\it Blue and red symbols}  show the integrated spectra of super-soft X-ray sources, i.e. sources located to the left of and above the stable nuclear burning boundary in late-type and early-type galaxies respectively. The spectra are normalized to the same flux at 0.35 keV. See Section \ref{sec:SSS_sample} for more detail.
}
\label{fig:spectra}
\end{figure}

Along with rare super-soft X-ray sources, in all galaxies from our sample there are much more abundant populations of soft sources with somewhat harder, but still soft, spectra, located to the right  of the stable nuclear burning boundary in  Fig. \ref{fig:lum_kt}.  These sources were previously classified as quasi-soft X-ray sources \citep{2004ApJ...609..710D,2006ApJ...643..844O}. Among other hypothesis, it was proposed that they are associated with supernova remnants or with accreting black holes in the soft spectral state \citep[e.g.][]{2004ApJ...609..710D,2013ApJ...771....6P, 2014ApJS..212...21L, 2016ApJ...827...46K}, although the contribution of the latter should be minimal (Section  \ref{sec:temp})

The  X-ray emitting SNRs will be largely unresolved by Chandra and appear as point sources. Indeed, the SNR emission  shifts outside X-ray band when the shock slows down to about  $\sim 200$ km/sec and the post-shock  temperature drops below $\sim 10^6$ K \citep{2017hsn..book.2063V}. This happens about $\sim 2\times10^4$ years after the explosion when the  remnant size  typically reaches $\sim 10-20$ pc  \citep{blondin1998}. At the distances of spiral galaxies 3.6$\sim$7.6 Mpc, linear dimension of $10-20$ pc corresponds to the angular size of $\sim 0.4-0.9\arcsec$, which can be barely resolved with Chandra ACIS for the brightest sources. Furthermore, the majority of supernova remnants will be classified by hardness ratio and median energy as soft sources.

To clarify the nature of soft sources, located to the right from the stable nuclear burning boundary, we used M83 as a test case. It is a galaxy with one of the best studied X-ray SNR population.   \citet{2014ApJS..212...21L} identified 87 X-ray detected SNR  candidates based on their X-ray spectral properties, associations with optical emission nebulae and/or radio sources. From this sample, SNR nature of 67 sources was confirmed with optical spectroscopy using [SII]/H$\alpha$ line ratio \citep{2017ApJ...839...83W}. The list of 87 X-ray detected  SNRs from \citet{2014ApJS..212...21L} was cross-correlated with our catalog of X-ray sources detected in M83. With 1$\arcsec$ matching radius, we found  66 matches, of which 6 sources have statistically significant emission beyond 2 keV and remaining 60 sources were classified as soft sources in our analysis. We replot $kT_{\rm bb}-L_X$ diagramm for M83  in Fig. \ref{fig:m83} marking confirmed  SNRs by cyan symbols. As one can see, all confirmed SNRs are located outside the stable hydrogen nuclear burning boundary  on $kT_{\rm bb}-L_{\rm X}$ diagram (i.e. to the right of the thick red line in Fig. \ref{fig:m83}) and their distribution on the $kT_{\rm bb}-L_X$ plane is morphologically quite similar to the distribution of the bulk of quasi-soft sources. We therefore conclude that the majority of quasi-soft sources must be SNRs.

For many of the brightest quasi-soft sources  in spiral galaxies likelihood values obtained in spectral fitting suggest that the spectrum is more complicated than predicted by a featureless black body model. A more detailed investigation of several such  sources revealed  that these deviations are mostly associated with the presence of emission lines in their spectra. In majority of cases the spectrum of thermal emission of optically thin plasma or a non-equilibrium ionization model significantly improved the quality of the fit, suggesting  that  these sources may be SNRs. However, such a detailed analysis is only possible for a handful of brightest sources.

 In order to compare spectral properties of soft and super-soft sources, we combined their spectra in late- and early- type galaxies. Combined spectra are shown in  Fig.\ref{fig:spectra} separately for sources located inside and outside the stable hydrogen nuclear burning boundary on the $kT_{bb}-L_X$ diagram (see Section \ref{sec:temp}). For comparison, we also show the combined spectrum of confirmed SNRs in M83. As one can see, combined spectrum of soft sources in spiral galaxies shows strong Mg and S lines and matches very well the combined spectrum of SNRs in M83. Similar values of the equivalent width of Mg and S lines support our conclusion that the vast majority of soft sources in these galaxies must be supernova remnants.
In early-type galaxies, on the contrary, combined spectrum of soft sources is harder and does not show strong emission lines.  As it was discussed in Section \ref{sec:temp}, by the method of selection, soft sources in early-type  galaxies should include large numbers of  X-ray binaries and background AGN. On the other hand, super-soft sources in all galaxies have much softer spectra, as expected. Their spectra are broadly similar  in late- and early-type galaxies, although the quality of the latter does not permit to make a more quantitative conclusion.

We finally note that the numbers and specific frequencies of soft sources in galaxies from our sample seem to follow same trend as super-soft sources, decreasing from late-type to early-type galaxies. However a detailed investigation of dependence of the specific frequency of SNRs on the morphological type would require a more accurate selection of sources and is outside the scope of the present paper.

\section{Summary}
\label{sec:summary}

Based on  public {\it Chandra} archival data, we studied populations of soft and super-soft X-ray sources with the goal to  investigate populations of stable  nuclear  burning accreting WDs.

To this end, we searched Chandra data archive  and constructed a sample of galaxies optimal for studying populations of steady nuclear burning accreting white dwarfs. We aimed at constructing the best possible sample of such galaxies. We thus selected   four late-type and five early-type galaxies with stellar masses $\ga$ $10^{10}$ $M_\odot$ and Chandra X-ray sensitivity limits comparable or better than  $\sim$ 10$^{36}$ ergs/s.  Our selection presents the nearly full sample of  galaxies available in Chandra archive up to cycle 20, suitable for such a study (among galaxies fitting within the field of view of Chandra observation)

We proposed a new approach to identify  population of soft and super-soft  X-ray sources based on combination of hardness ratio and median energy.
Distributions of sources over median energy in late-type galaxies shows clear bi-modality with a distinct populations of soft and super-soft
sources. These sources  show a rather compact  distributions in the X-ray luminosity--temperature plane with temperatures of 20--300 eV and X-ray luminosities in the 10$^{36}$$\sim$10$^{38}$ ergs/s range. The picture is more complex in lenticular and elliptical galaxies, where populations of soft sources smoothly connect to the populations of X-ray binaries and background AGN in their median energy distributions. For these galaxies we mainly used the lack of emission above 2 keV to select soft sources for detailed analysis.

We used X-ray luminosity--temperature diagram to further classify selected sources into super-soft and soft. To this end we used the boundary of stable hydrogen nuclear burning to separate super-soft X-ray sources -- candidates to nuclear burning white dwarfs from the rest of the population of soft sources. In late-type galaxies, we found 29 super-soft source which yields the specific frequency of  $f_{\rm SSS}^{\rm S}$$\approx$(2.08$\pm$0.46)$\times10^{-10}$ M$_{\odot}^{-1}$. In   lenticular and elliptical galaxies,  there are  7 super-soft sources with the  specific frequency of $f_{\rm SSS}^{\rm E+S0}$$\approx$ (2.47$\pm$1.34)$\times10^{-11}$ M$_{\odot}^{-1}$. Specific frequencies in late-type and early-type galaxies differ at the statistical confidence of  $\approx 5.8 \sigma$. The ratio of specific frequencies of super-soft X-ray sources in late-type galaxies and early-type galaxies is $f_{\rm SSS}^{\rm E+S0}/ f_{\rm SSS}^{\rm S}=0.12\pm0.05$.  Our results are broadly consistent with the population synthesis modelling of the populations  of stable nuclear burning accreting WDs.

\section{Acknowledgements}
This research has made use of data obtained from the Chandra Data Archive and software provided by the Chandra X-ray Center (CXC) in the application packages CIAO. This publication makes use of data products from the HyperLeda and RC3 database. This publication makes use of data products from the Two Micron All Sky Survey, which is a joint project of the University of Massachusetts and the Infrared Processing and Analysis Center/California Institute of Technology, funded by the National Aeronautics and Space Administration and the National Science Foundation; the NASA/IPAC Extragalactic Database (NED), which is operated by the Jet Propulsion Laboratory, California Institute of Technology, under contract with the National Aeronautics and Space Administration. The authors acknowledge partial support of this work by the Russian Government Program of Competitive Growth of Kazan Federal University (IG) and by the RSF grant 19-12-00369 (MG). The authors thank the anonymous referee for useful and inspiring suggestions which helped to improve the manuscript.

\bibpunct{(}{)}{;}{a}{}{,}
\bibliography{Bibliography.bib}

\begin{appendix} 

\section{List of observations}
\label{appendix:long_obs}
Table \ref{tab:obs_app} shows list of  {\it Chandra} observations used in this paper. Columns are: (1) -- Galaxy name; (2) -- Observation identification number; (3) -- Date of observations; (4) -- Exposure time; (5) -- Type of ACIS camera used during observations.

\longtab[1]{
\begin{longtable}{lllrc}
\caption{\label{tab:obs_app} List of  {\it Chandra} observations used for the analysis. } \\
\hline
Galaxy                  &ObsID                   & Date                                & Exposure (ksec)  & Camera      \\
                     (1)    &                                      (2)   & (3)      & (4)  & (5) \\
\hline
NGC 5236 (M83)    &12992  & 04/09/2011      & 66.3    & ACIS-S   \\
			      &12993  & 15/03/2011      & 49.4    & ACIS-S   \\
			      &12994  & 23/03/2011      & 150.1  & ACIS-S   \\
			      &12995  & 23/12/2010      & 59.3    & ACIS-S   \\
			      &12996  & 29/03/2011      & 53.0    & ACIS-S   \\
			      &13202  & 25/12/2010      & 98.8    & ACIS-S   \\
			      &13241  & 18/03/2011      & 79.0    & ACIS-S   \\
			      &13248  & 18/03/2011      & 54.3    & ACIS-S   \\
			      &14332  & 29/08/2011      & 52.4    & ACIS-S   \\
			      &14342  & 28/12/2011      & 67.1   & ACIS-S   \\
\hline
NGC 5194 (M51)    &13812  & 12/09/2012      &157.5    & ACIS-S   \\
			      &13813  & 09/09/2012      &179.2   & ACIS-S   \\
			      &13814  & 20/09/2012      &189.8   & ACIS-S   \\
			      &13815  & 23/09/2012      &67.2   & ACIS-S   \\
			      &13816  & 26/09/2012      &73.1   & ACIS-S   \\
			      &15496  & 19/09/2012      &41.0   & ACIS-S   \\
			      &15553  & 10/10/2012      &37.6   & ACIS-S   \\
\hline
NGC 3031 (M81)   &5935    & 26/05/2005      &11.0   &ACIS-S \\
			     &5936    & 28/05/2005      &11.4   &ACIS-S \\
       			     &5937    & 01/06/2005      &12.0   &ACIS-S \\
       			     &5938    &03/06/2005       &11.8   &ACIS-S \\
       			     &5939    &06/062005        &11.8   &ACIS-S \\
       			     &5940    &09/06/2005       &12.0  &ACIS-S \\
       			     &5941    &11/06/2005       &11.8   &ACIS-S \\
       			     &5942    &15/06/2005       &11.9   &ACIS-S \\
       			     &5943    &18/06/2005       &12.0   &ACIS-S \\
       			     &5944    &21/06/2005       &11.8   &ACIS-S \\
       			     &5945    &24/06/2005      &11.6   &ACIS-S \\
       			     &5946    &26/06/2005      &12.0   &ACIS-S \\
       			     &5947    &29/06/2005      &10.7   &ACIS-S \\
       			     &5948    &03/07/2005      &12.0   &ACIS-S \\
       			     &5949    &06/07/2005      &12.0   &ACIS-S \\
\hline
NGC 5457 (M101) &4731    &19/01/2004      &56.2   &ACIS-S \\
			     &4732    &19/03/2004      &69.8   &ACIS-S \\
			     &4733    &07/05/2004      &24.8   &ACIS-S \\
			     &4734    &11/07/2004      &35.5   &ACIS-S \\
			     &4735    &12/09/2004      &28.8   &ACIS-S \\
			     &4736    &01/11/2004      &77.3   &ACIS-S \\
			     &4737    &01/01/2005      &21.9   &ACIS-S \\
			     &5296    &21/01/2004      &3.1     &ACIS-S \\
			     &5297    &24/01/2004      &21.7   &ACIS-S \\
 			     &5300    &07/03/2004      &52.1   &ACIS-S \\
			     &5309    &14/03/2004      &70.8   &ACIS-S \\
			     &5322    &03/05/2004      &64.7   &ACIS-S \\
			     &5323    &09/05/2004      &42.6   &ACIS-S \\
			     &5337    &05/07/2004      &9.9     &ACIS-S \\
			     &5338    &06/07/2004      &28.6   &ACIS-S \\
			     &5339    &07/07/2004      &14.3   &ACIS-S \\
			     &5340    &08/07/2004      &54.4   &ACIS-S \\
			     &6114    &05/09/2004      &66.2   &ACIS-S \\
			     &6115    &08/09/2004      &35.8   &ACIS-S \\
			     &6118    &11/09/2004      &11.5   &ACIS-S \\
			     &6152    &07/11/2004      &44.1   &ACIS-S \\
			     &6169    &30/12/2004      &29.4   &ACIS-S \\
			     &6170    &22/12/2004      &48.0   &ACIS-S \\
			     &6175    &24/12/2004      &40.7   &ACIS-S \\
\hline
NGC 5128 (Cen A) &7797   &22/03/2007      &96.9   &ACIS-I \\
			     &7798   &27/03/2007      &90.8   &ACIS-I \\
			     &7799   &30/03/2007      &94.8   &ACIS-I \\
			     &7800   &17/04/2007      &90.8   &ACIS-I \\
			     &8489   &08/05/2007      &93.9   &ACIS-I \\
			     &8490   &30/05/2007      &94.4   &ACIS-I \\
\hline
NGC 3379   	      &7073   &23/01/2006      &84.1   &ACIS-S\\
			      &7074   &09/04/2006      &69.1   &ACIS-S\\
			      &7075   &07/03/2006      &83.1   &ACIS-S\\
			      &7076   &10/01/2007      &69.2   &ACIS-S\\
\hline
NGC 3115              &13817  &18/01/2012      &172.0     &ACIS-S\\
			      &13819  &26/01/2012     &75.5       &ACIS-S\\
			      &13820  &31/01/2012     &184.2     &ACIS-S\\
			      &13821  &03/02/2012     &158.0     &ACIS-S\\
			      &13822  &21/01/2012     &160.2     &ACIS-S\\
			      &14383  &04/04/2012     &119.5     &ACIS-S\\
			      &14384  &06/04/2012     &69.7       &ACIS-S\\
			      &14419  &05/04/2012     &46.3       &ACIS-S\\
\hline
NGC 4278              &7077  &16/03/2006      &110.3     &ACIS-S\\
			      &7078  &25/07/2006      &51.4      &ACIS-S\\
			      &7079  &24/10/2006      &105.1    &ACIS-S\\
			      &7080  &20/04/2007      &55.8      &ACIS-S\\
			      &7081  &20/02/2007      &110.7     &ACIS-S\\
\hline
NGC 4697              &4727  &26/12/2003      &39.9       &ACIS-S\\
			      &4728  &06/01/2004      &35.7      &ACIS-S\\
			      &4729  &12/02/2004      &38.1      &ACIS-S\\
			      &4730  &18/08/2007      &40.0      &ACIS-S\\
\hline
\end{longtable}
}

\section{Galaxy images}
\label{appendix:image}

Fig. \ref{fig:galaxy_images} shows X-ray  images of galaxies in our sample and outlines of analysed regions.
\begin{figure*}
\begin{center}
\hbox
{
\includegraphics[width=60mm]{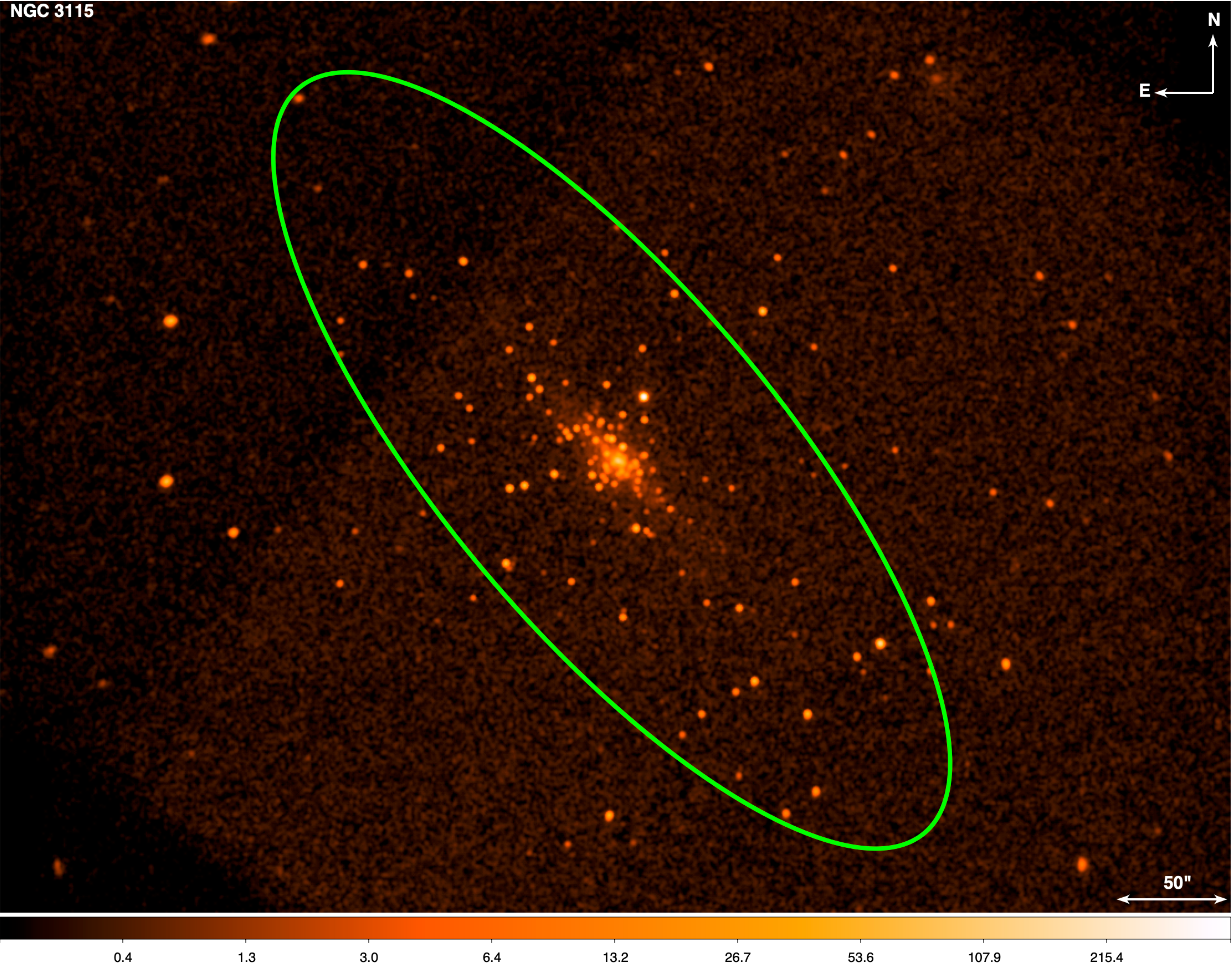}
\includegraphics[width=60mm]{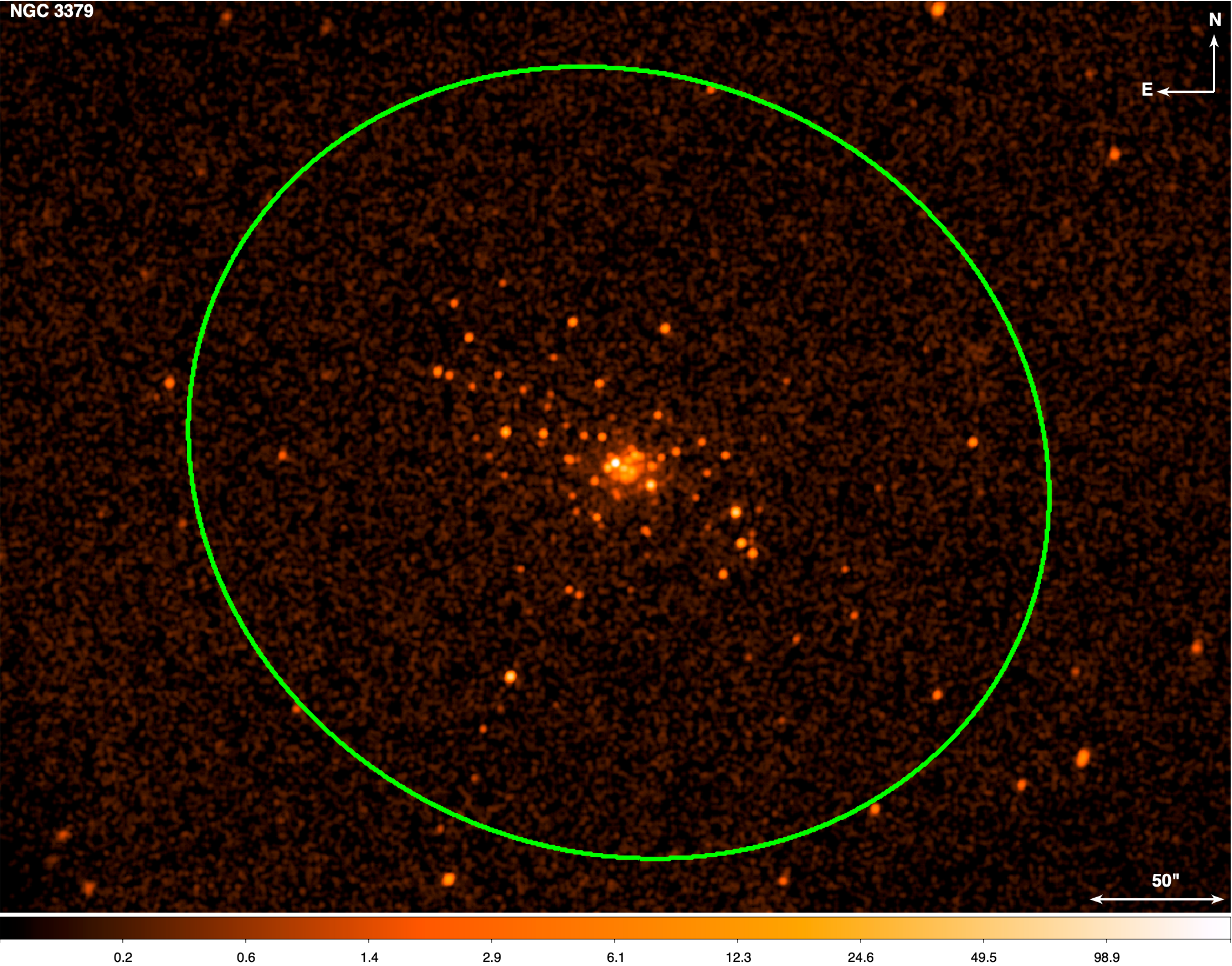}
\includegraphics[width=60mm]{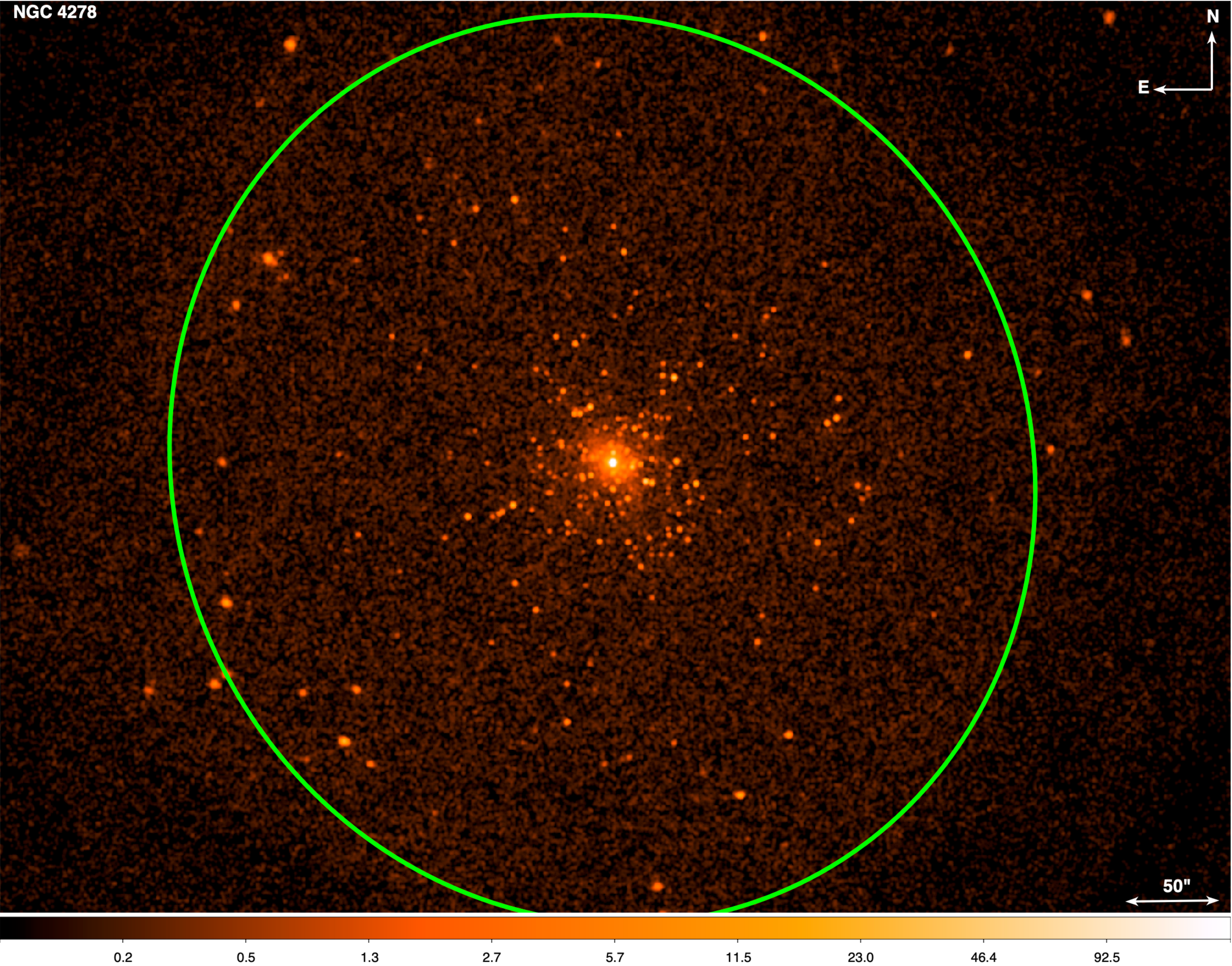}
}
\hbox
{
\includegraphics[width=60mm]{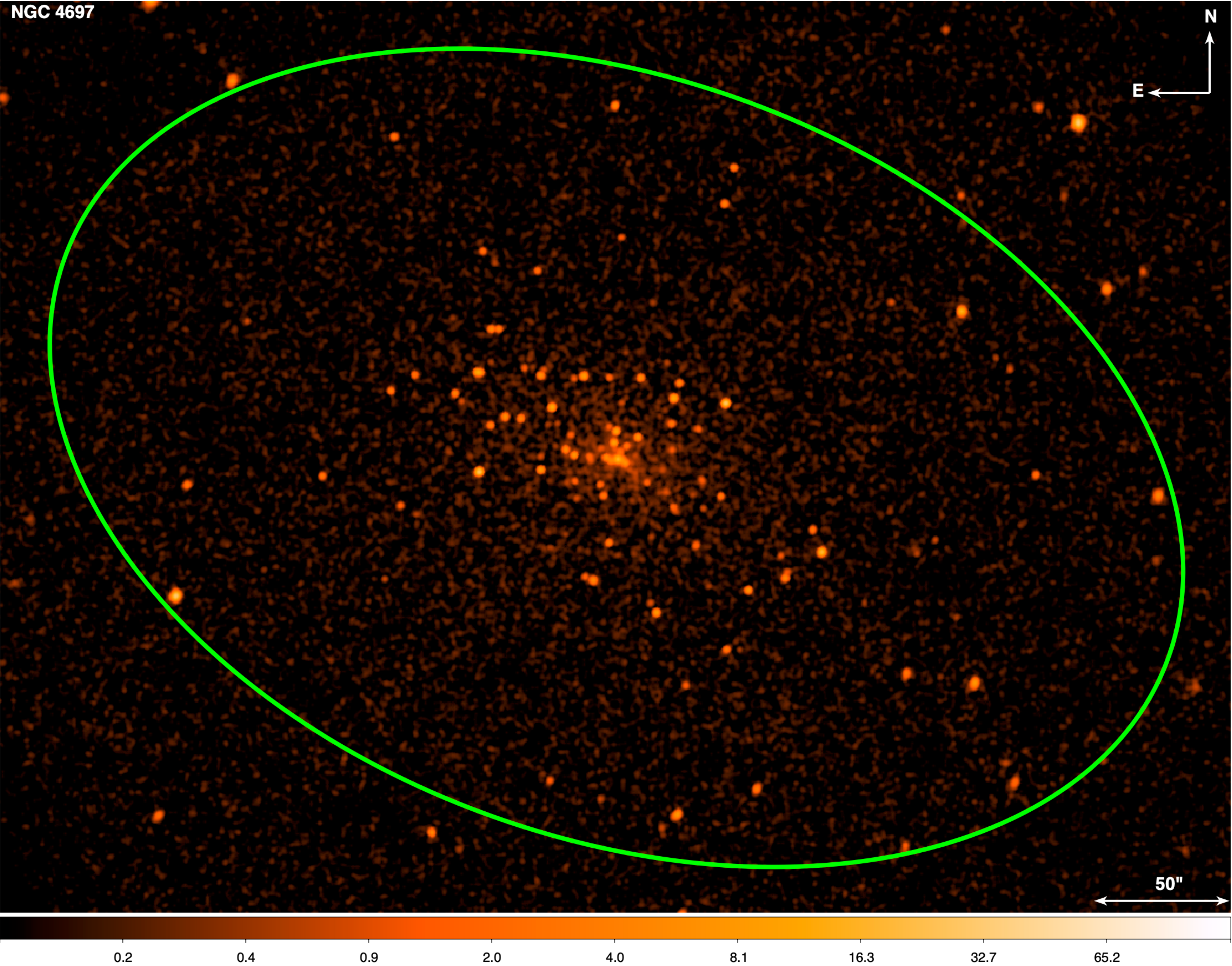}
\includegraphics[width=60mm]{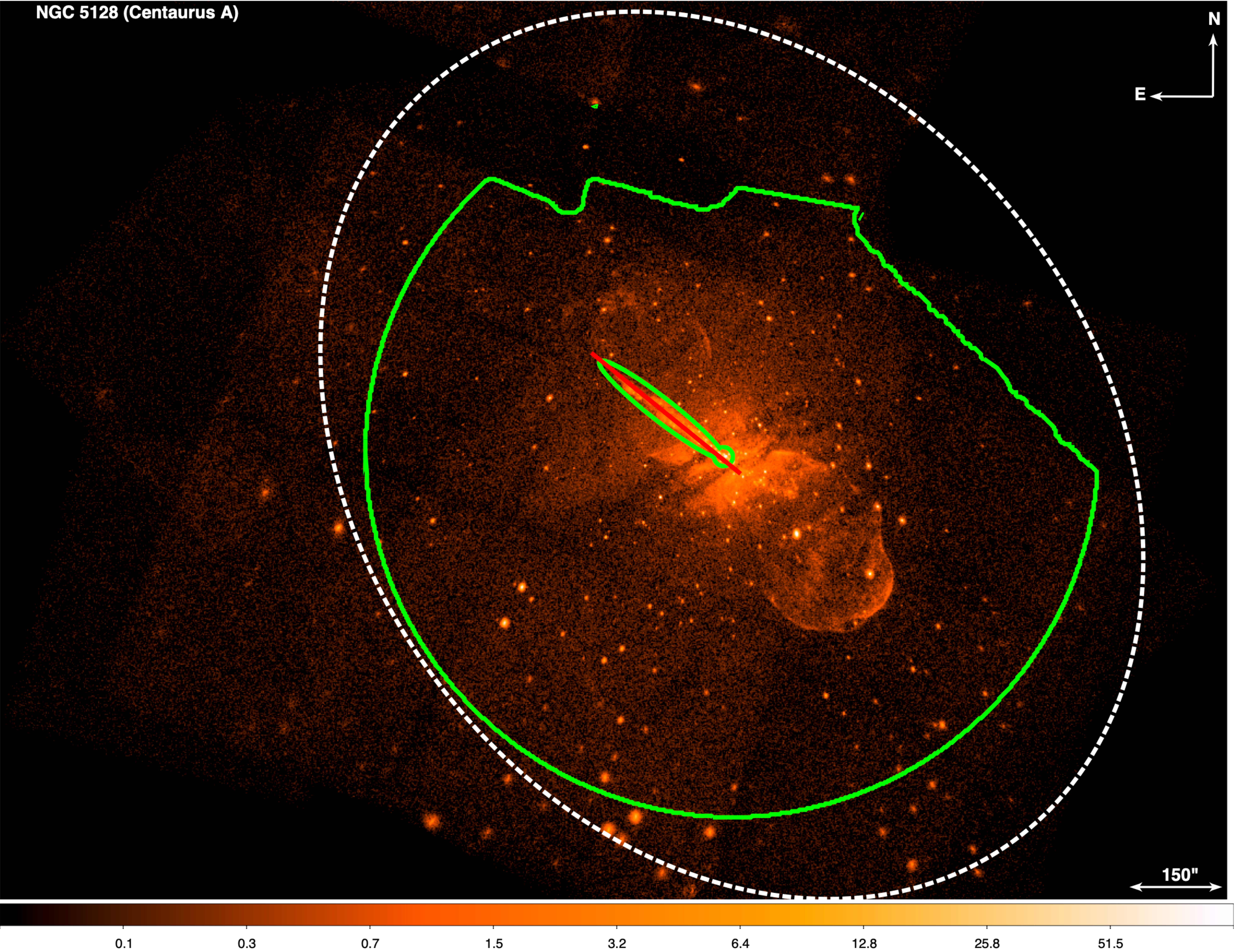}
\includegraphics[width=60mm]{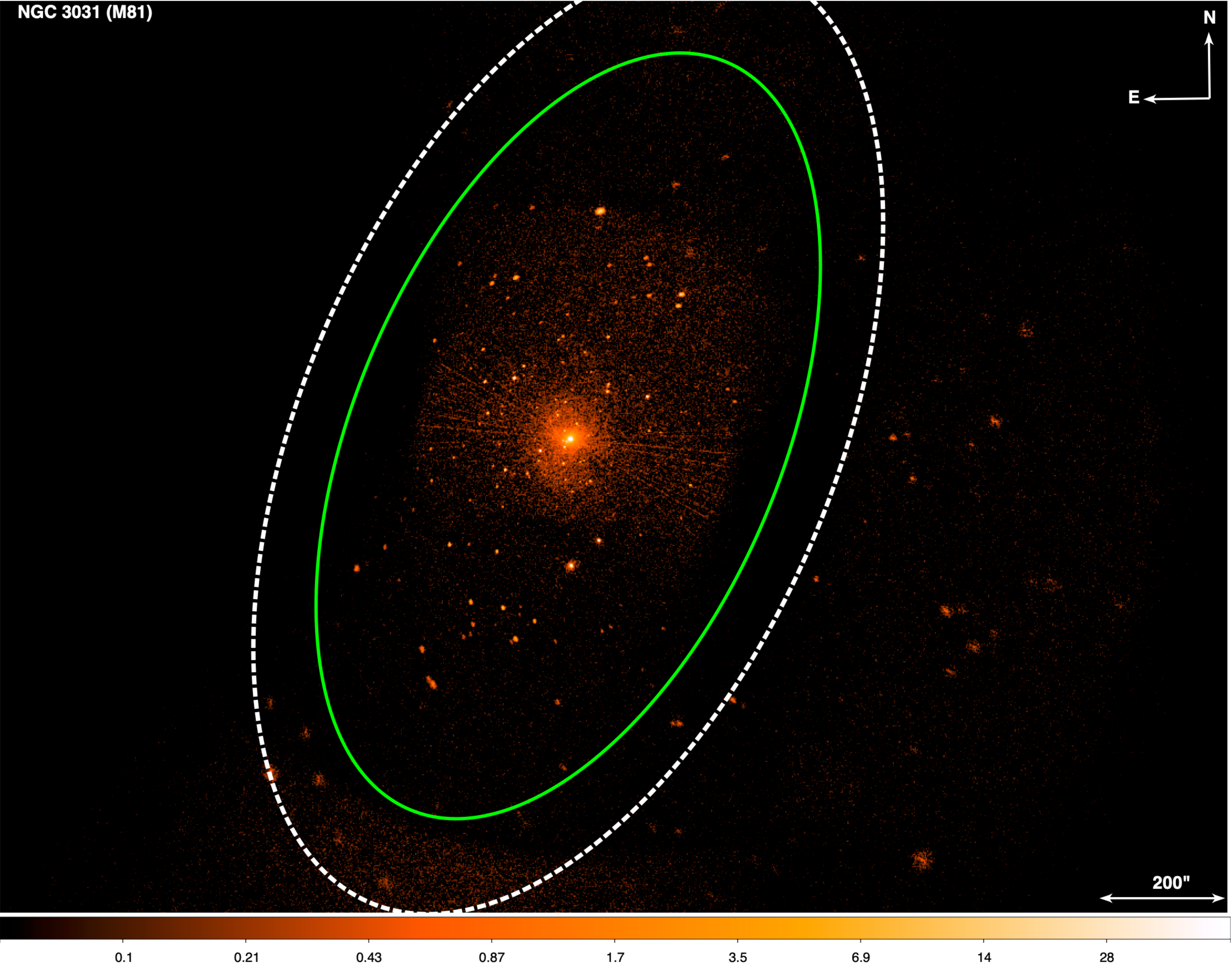}
}
\hbox
{
\includegraphics[width=60mm]{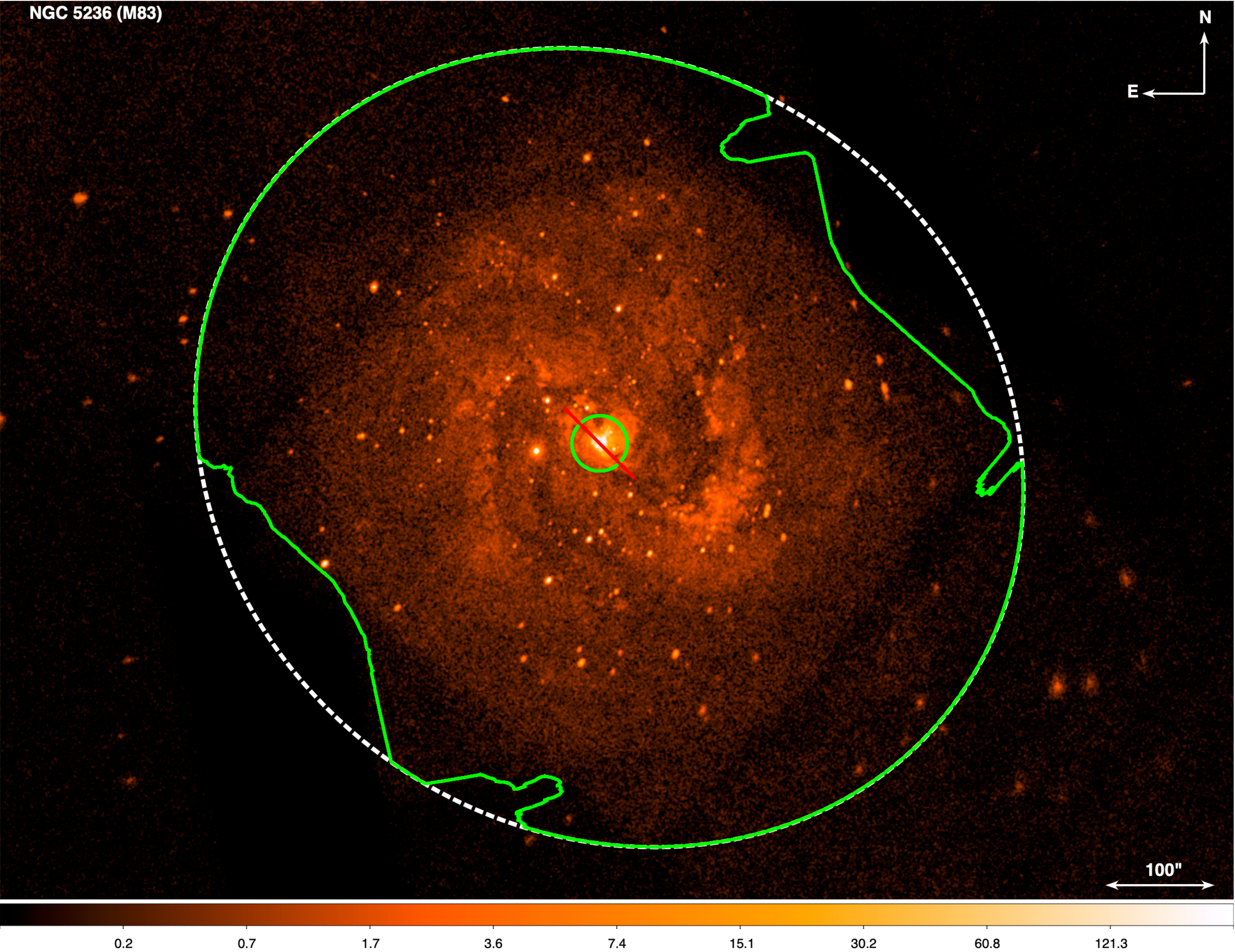}
\includegraphics[width=60mm]{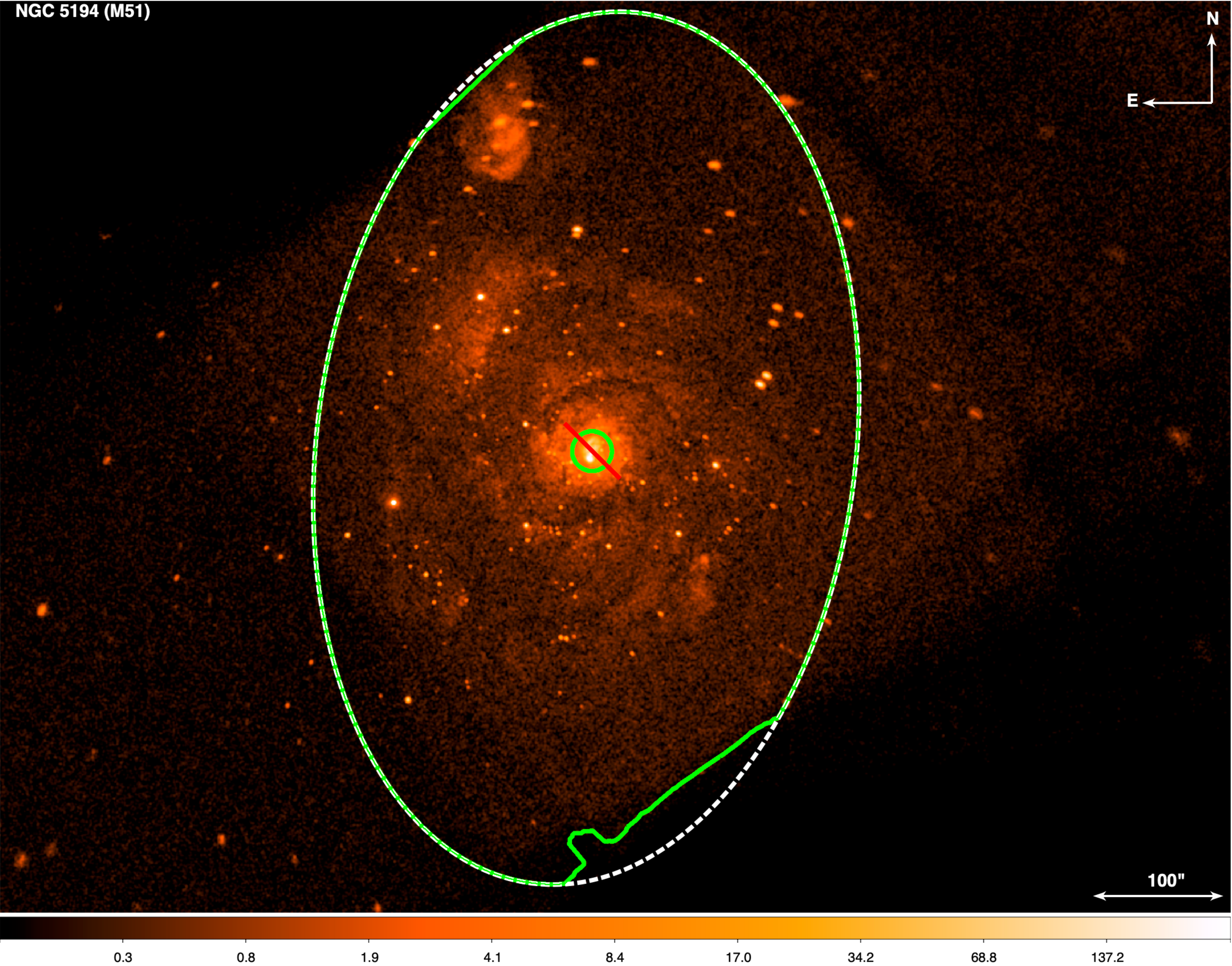}
\includegraphics[width=60mm]{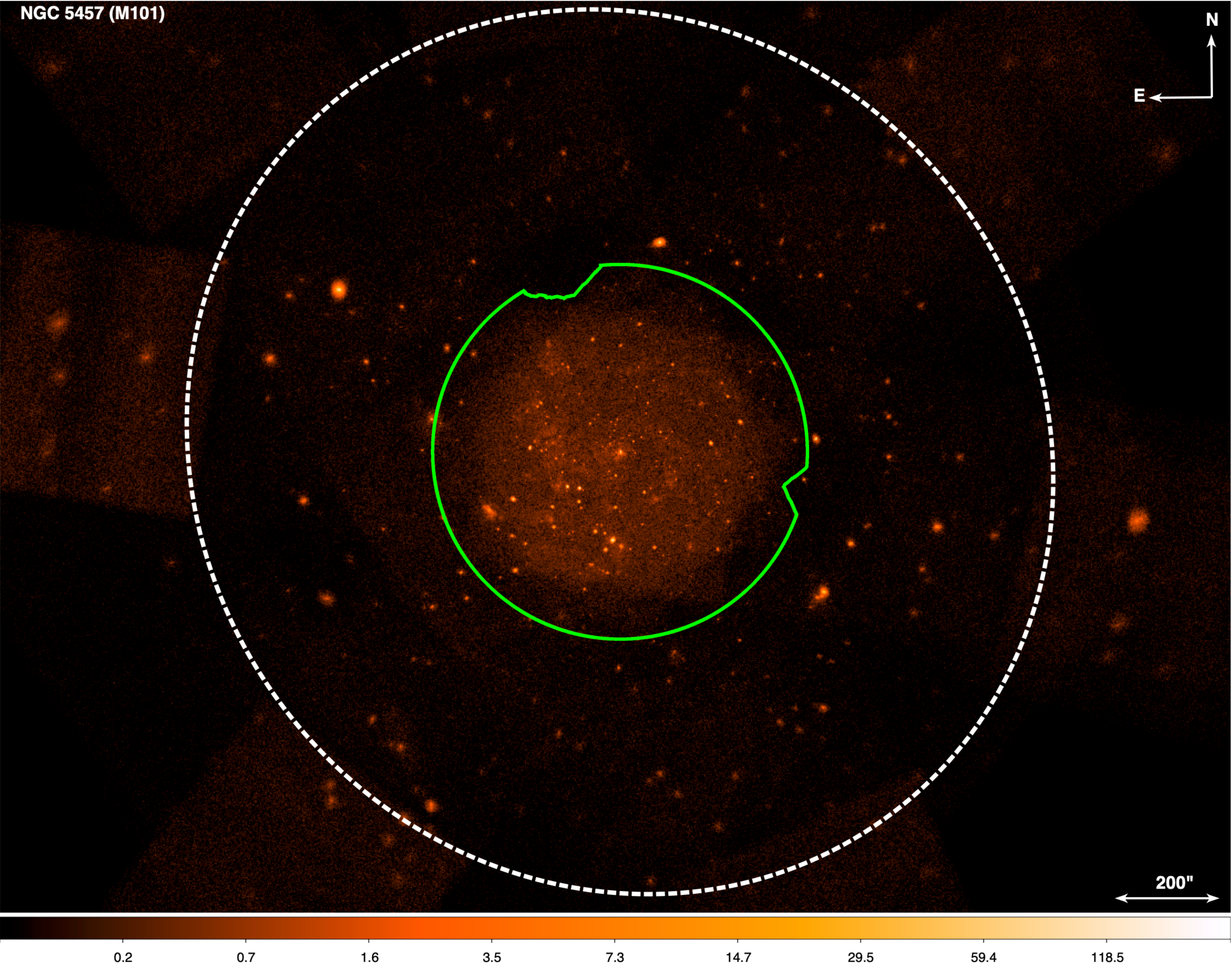}
}
\caption{ The 0.3--2 keV band false-color images (smoothed)  of  galaxies in our sample. Dashed white regions correspond to D$_{25}$. Green lines show regions used for source detection having exposure map values greater than 20 $\%$ of maximal value (for more detail see Section \ref{sec:src_det}). Stellar masses listed in Table 2 are computed for these regions. The inside areas of small green regions near the center of Centaurus A, M83 and  M51 were excluded from analysis.}
\label{fig:galaxy_images}
\end{center}
\end{figure*}

\section{Catalog of super-soft X-ray sources}
\label{appendix:catalog}

The catalog is comprised of super-soft X-ray sources with more than 20 counts (solid black line in Fig. \ref{fig:lum_kt}.)  The sources are classified as super-soft if they are located to the left of the stable nuclear burning boundary (nearly vertical solid red line in n Fig. \ref{fig:lum_kt}) or their error bars cross this boundary.
We also included two sources in M101, located outside but close to the stable nuclear burning boundary (the two sources in Fig.\ref{fig:lum_kt} upper right panel with $kT\ga 100$ eV and $L_X\ga 10^{37}$ erg/s.). The catalog of super-soft X-ray sources in given in Table \ref{table:catalog}.

\longtab[1]{
\begin{landscape}
\renewcommand{\arraystretch}{1.3}
\begin{longtable}{l l l l r r r r r r}
\caption{\label{table:catalog} List of super-soft X-ray sources.} \\
\hline
ID & GALAXY & R.A. & DEC. & Counts$_{0.3-2}$ & Bkg$_{0.3-2}$ & Rate$_{0.3-2}$ & N$_{\rm H}$ & kT$_{\rm bb}$ & L$_{\rm X}^{0.3-2}$ \\
    &     &     &     &     &     & (10$^{-4}$ cnt/s) & (10$^{21}$cm$^{-2}$) & (eV) & (10$^{37}$ erg/s)  \\
(1) & (2) & (3) & (4) & (5) & (6) & (7) & (8) & (9) & (10)  \\
\hline
1 ....... & M81 & 09 55 42.14& +69 03 36.40&     2565.0 $\pm$  51.7&   8.7 $\pm$   0.9&     145.30 $\pm$  2.94&  1.19&   85$^{+    2}_{-    1}$&     15.87$^{+ 0.34}_{- 0.32}$ \\
2 ....... & M81 & 09 56  8.98& +69 01  6.79& 360.0 $\pm$  20.0&   1.6 $\pm$   0.5& 20.37 $\pm$  1.14&  1.39&   86$^{+    4}_{-    3}$& 7.04$^{+ 0.48}_{- 0.46}$ \\
3 ....... & NGC3379 & 10 47 47.23& +12 34 59.78&  72.0 $\pm$   9.5&   0.5 $\pm$   0.3&  3.03 $\pm$  0.40&  0.26&  100$^{+   18}_{-   13}$& 3.68$^{+ 0.46}_{- 0.42}$ \\
4 ....... & NGC4278 & 12 19 56.67& +29 16 31.08&  20.0 $\pm$   5.6&   1.6 $\pm$   0.5&  0.42 $\pm$  0.13&  0.20&   65$^{+   18}_{-   13}$& 2.76$^{+ 0.82}_{- 0.63}$ \\
5 ....... & NGC4278 & 12 20 12.43& +29 17 41.86&  25.0 $\pm$   6.1&   1.3 $\pm$   0.4&  0.55 $\pm$  0.14&  0.20&   72$^{+   18}_{-   13}$& 2.26$^{+ 0.64}_{- 0.50}$ \\
6 ....... & NGC4697 & 12 48 34.55& --5 47 49.34&  36.0 $\pm$   7.1&   0.8 $\pm$   0.3&  2.29 $\pm$  0.46&  0.21&  102$^{+   24}_{-   15}$& 2.79$^{+ 0.58}_{- 0.47}$ \\
7 ....... & NGC4697 & 12 48 41.28& --5 48 19.73&  24.0 $\pm$   6.0&   0.2 $\pm$   0.2&  1.55 $\pm$  0.39&  0.21&   70$^{+   23}_{-   12}$& 2.41$^{+ 0.63}_{- 0.52}$ \\
8 ....... & CenA & 13 25 14.60& --42 56 11.57&  83.0 $\pm$  10.2&  19.3 $\pm$   1.4&  2.28 $\pm$  0.37&  0.24&   79$^{+    9}_{-    8}$& 2.60$^{+ 0.56}_{- 0.47}$ \\
9 ....... & CenA & 13 25 18.14& --43 03 42.53& 100.0 $\pm$  11.0&  27.8 $\pm$   1.7&  1.29 $\pm$  0.20&  0.24&   73$^{+    7}_{-    6}$& 2.06$^{+ 0.43}_{- 0.36}$ \\
10 ....... & M51 & 13 29 41.23& +47 11 15.57&  48.0 $\pm$   8.0&  16.1 $\pm$   1.2&  0.43 $\pm$  0.11&  0.92&   33$^{+   20}_{-    8}$& 1.27$^{+ 0.56}_{- 0.39}$ \\
11 ....... & M51 & 13 29 46.00& +47 10 56.36&  55.0 $\pm$   8.5&   9.5 $\pm$   1.0&  0.61 $\pm$  0.11&  0.69&   67$^{+   13}_{-   13}$& 0.96$^{+ 0.19}_{- 0.16}$ \\
12 ....... & M51 & 13 29 51.69& +47 12  3.82&  86.1 $\pm$  10.3&  56.9 $\pm$   2.3&  0.39 $\pm$  0.14&  3.93&   62$^{+   28}_{-   32}$& 0.29$^{+ 0.09}_{- 0.10}$ \\
13 ....... & M51 & 13 29 54.02& +47 11 25.68&  64.0 $\pm$   9.0&  15.4 $\pm$   1.2&  0.65 $\pm$  0.12&  2.89&   40$^{+   12}_{-   16}$& 0.57$^{+ 0.12}_{- 0.12}$ \\
14 ....... & M51 & 13 29 54.99& +47 12  4.74&  62.0 $\pm$   8.9&  16.0 $\pm$   1.3&  0.62 $\pm$  0.12&  2.92&   76$^{+   14}_{-   12}$& 0.53$^{+ 0.09}_{- 0.09}$ \\
15 ....... & M83 & 13 36 52.45& --29 52 52.09& 242.0 $\pm$  16.6& 113.3 $\pm$   3.2&  1.76 $\pm$  0.23&  5.40&   53$^{+    7}_{-    8}$& 0.42$^{+ 0.05}_{- 0.05}$ \\
16 ....... & M83 & 13 36 52.66& --29 49 37.45&  67.0 $\pm$   9.2&   9.1 $\pm$   1.0&  0.79 $\pm$  0.13&  1.32&   35$^{+    4}_{-    6}$& 0.87$^{+ 0.18}_{- 0.15}$ \\
17 ....... & M83 & 13 36 53.97& --29 50 32.52&  40.0 $\pm$   7.4&  14.3 $\pm$   1.2&  0.35 $\pm$  0.10&  3.60&   33$^{+    6}_{-    9}$& 0.19$^{+ 0.04}_{- 0.05}$ \\
18 ....... & M83 & 13 36 56.45& --29 52 57.25&  50.0 $\pm$   8.1&  21.3 $\pm$   1.5&  0.39 $\pm$  0.11&  5.00&   26$^{+    4}_{-    7}$& 0.20$^{+ 0.04}_{- 0.05}$ \\
19 ....... & M83 & 13 36 56.77& --29 53 16.20&  95.0 $\pm$  10.8&  22.5 $\pm$   1.5&  0.99 $\pm$  0.15&  3.39&   46$^{+    6}_{-    6}$& 0.35$^{+ 0.05}_{- 0.05}$ \\
20 ....... & M83 & 13 36 58.62& --29 51 56.84&  88.0 $\pm$  10.4&  49.2 $\pm$   2.1&  0.53 $\pm$  0.15& 10.41&   74$^{+   15}_{-   22}$& 0.11$^{+ 0.02}_{- 0.03}$ \\
21 ....... & M83 & 13 36 58.64& --29 53 38.86&  87.0 $\pm$  10.4&  30.6 $\pm$   1.7&  0.77 $\pm$  0.14&  1.82&   30$^{+    5}_{-    7}$& 0.69$^{+ 0.15}_{- 0.13}$ \\
22 ....... & M83 & 13 36 59.04& --29 52 18.07&  50.0 $\pm$   8.1&  22.5 $\pm$   1.5&  0.38 $\pm$  0.11& 11.69&   71$^{+   12}_{-   18}$& 0.09$^{+ 0.02}_{- 0.02}$ \\
23 ....... & M83 & 13 36 59.25& --29 51 43.82&  85.0 $\pm$  10.3&  49.4 $\pm$   2.2&  0.49 $\pm$  0.14&  9.95&   69$^{+   12}_{-   20}$& 0.11$^{+ 0.02}_{- 0.03}$ \\
24 ....... & M83 & 13 36 59.34& --29 53 17.90&  49.0 $\pm$   8.1&  19.1 $\pm$   1.3&  0.41 $\pm$  0.11&  2.50&   65$^{+   15}_{-   18}$& 0.12$^{+ 0.03}_{- 0.03}$ \\
25 ....... & M83 & 13 37  0.45& --29 50 54.02& 173.0 $\pm$  14.2&   9.1 $\pm$   1.0&  2.25 $\pm$  0.19&  4.05&   28$^{+    3}_{-    3}$& 1.13$^{+ 0.10}_{- 0.10}$ \\
26 ....... & M83 & 13 37  3.86& --29 52 23.02&  41.0 $\pm$   7.5&  10.8 $\pm$   1.0&  0.41 $\pm$  0.10&  4.36&   42$^{+    7}_{-    8}$& 0.18$^{+ 0.03}_{- 0.04}$ \\
27 ....... & M83 & 13 37  5.53& --29 50 32.24&  34.0 $\pm$   6.9&   6.0 $\pm$   0.8&  0.42 $\pm$  0.10&  3.01&   42$^{+    8}_{-   13}$& 0.21$^{+ 0.05}_{- 0.05}$ \\
28 ....... & M83 & 13 37 18.55& --29 52  7.59&  29.0 $\pm$   6.5&  10.6 $\pm$   1.0&  0.25 $\pm$  0.09&  1.57&   71$^{+   16}_{-   23}$& 0.12$^{+ 0.04}_{- 0.03}$ \\
29 ....... & M101 & 14 02 51.34& +54 19 18.24& 178.0 $\pm$  14.4&  14.7 $\pm$   1.2&  2.02 $\pm$  0.18&  1.50&   85$^{+    6}_{-    5}$& 0.91$^{+ 0.08}_{- 0.07}$ \\
30 ....... & M101 & 14 02 51.64& +54 22  3.97&  33.0 $\pm$   6.8&  10.5 $\pm$   1.0&  0.28 $\pm$  0.08&  1.17&   70$^{+   24}_{-   17}$& 0.13$^{+ 0.03}_{- 0.03}$ \\
31 ....... & M101 & 14 03  1.22& +54 23 41.41& 324.0 $\pm$  19.0&  13.3 $\pm$   1.1&  3.83 $\pm$  0.23&  1.13&   66$^{+    3}_{-    3}$& 1.85$^{+ 0.12}_{- 0.11}$ \\
32 ....... & M101 & 14 03 12.75& +54 21 11.52&  38.0 $\pm$   7.2&  13.1 $\pm$   1.1&  0.28 $\pm$  0.08&  1.99&   73$^{+   18}_{-   14}$& 0.10$^{+ 0.02}_{- 0.02}$ \\
33 ....... & M101 & 14 03 13.65& +54 20  9.39& 700.0 $\pm$  27.5&  10.9 $\pm$   1.0&  7.24 $\pm$  0.29&  1.57&   53$^{+    2}_{-    2}$& 3.17$^{+ 0.13}_{- 0.13}$ \\
34 ....... & M101 & 14 03 15.52& +54 17  3.81& 339.0 $\pm$  19.4&  30.7 $\pm$   1.7&  3.24 $\pm$  0.20&  2.52&  104$^{+    5}_{-    5}$& 1.24$^{+ 0.07}_{- 0.07}$ \\
35 ....... & M101 & 14 03 16.52& +54 20 54.80&  44.0 $\pm$   7.7&  12.5 $\pm$   1.1&  0.34 $\pm$  0.08&  1.85&   66$^{+   14}_{-   11}$& 0.16$^{+ 0.03}_{- 0.03}$ \\
36 ....... & M101 & 14 03 19.02& +54 17 19.48& 254.0 $\pm$  17.0&  33.5 $\pm$   1.8&  2.32 $\pm$  0.18&  1.40&   61$^{+    4}_{-    4}$& 1.32$^{+ 0.10}_{- 0.09}$ \\
37 ....... & M101 & 14 03 27.38& +54 21 11.62& 109.0 $\pm$  11.5&  24.4 $\pm$   1.5&  0.89 $\pm$  0.12&  1.67&   27$^{+    4}_{-    3}$& 0.83$^{+ 0.12}_{- 0.11}$ \\
38 ....... & M101 & 14 03 29.91& +54 20 57.27& 259.0 $\pm$  17.1&  16.7 $\pm$   1.3&  2.54 $\pm$  0.18&  1.81&   64$^{+    4}_{-    3}$& 1.15$^{+ 0.08}_{- 0.08}$ \\
39 ....... & M101 & 14 03 33.35& +54 17 59.73& 433.0 $\pm$  21.8&  22.2 $\pm$   1.4&  4.32 $\pm$  0.23&  2.24&   60$^{+    3}_{-    2}$& 2.62$^{+ 0.14}_{- 0.14}$ \\
\hline
\end{longtable}
Description of columns: (1)  Source ID; (2) Galaxy name; (3) Right ascension (J2000); (4) Declination (J2000); (5) Counts in the source region  in the 0.3-2 keV band and its 1$\sigma$ error; (6) Background counts in the source region  in 0.3--2 keV band and its 1$\sigma$ error (68\%); (7) Source count rate in 0.3--2 keV band ; (8) Absorption column density in units of 10$^{21}$ cm$^{-2}$ obtained by combining CO 2-1 and 21 cm  data as described in  Section \ref{sec:nh}; (9) best fit color temperature  in units of eV;  (10)   absorbed X-ray luminosity in the  0.3-2 keV band  in units of 10$^{37}$ erg s$^{-1}$. Parameters given in columns (9) and (10) were obtained by approximating spectra with the  black body model with absorption fixed at the value from column (8).
\end{landscape}
}

\end{appendix}
\end{document}